\documentstyle[12pt,aaspp]{article}

\begin{document}

\def\MSUN{\rm M_{\odot}}
\def\RSUN{\rm R_{\odot}} 
\def\MSUNYR{\rm M_{\odot}\,yr^{-1}}
\def\MDOT{\dot{M}}

\newbox\grsign \setbox\grsign=\hbox{$>$} \newdimen\grdimen \grdimen=\ht\grsign
\newbox\simlessbox \newbox\simgreatbox
\setbox\simgreatbox=\hbox{\raise.5ex\hbox{$>$}\llap
     {\lower.5ex\hbox{$\sim$}}}\ht1=\grdimen\dp1=0pt
\setbox\simlessbox=\hbox{\raise.5ex\hbox{$<$}\llap
     {\lower.5ex\hbox{$\sim$}}}\ht2=\grdimen\dp2=0pt
\def\simgreat{\mathrel{\copy\simgreatbox}}
\def\simless{\mathrel{\copy\simlessbox}}

\title{ Accretion of low angular momentum material onto black holes:
2D hydrodynamical inviscid case.
}

\vspace{1.cm}
\author{ Daniel Proga and Mitchell C. Begelman$^{1}$}
\vspace{.5cm}
\affil{JILA, University of Colorado, Boulder, CO 80309-0440, USA;
proga@colorado.edu, mitch@jila.colorado.edu}

$^1$ also Department of Astrophysical and Planetary Sciences, University
of Colorado at Boulder

\begin{abstract}
We report on the first phase of our study of slightly rotating accretion 
flows onto black holes. We consider inviscid accretion flows with 
a spherically symmetric density distribution at the outer boundary,
but with  spherical symmetry broken by the introduction of
a small, latitude-dependent angular momentum.
We study accretion flows by means of numerical 2D, axisymmetric, 
hydrodynamical simulations. Our  main result is that the properties of 
the accretion flow do not depend as much on the outer boundary conditions 
(i.e., the amount as well as distribution of the angular momentum) as on 
the geometry of the non-accreting matter. The material that has too much 
angular momentum to be accreted forms a thick torus near the equator. 
Consequently, the geometry of the polar region, where material is accreted 
(the funnel), and the mass accretion rate through it are constrained by 
the size and shape  of the torus. Our results show one way in which
the mass accretion 
rate of slightly rotating gas can be significantly reduced compared to 
the accretion of non-rotating gas (i.e., the Bondi rate), and set the stage
for calculations that will take into account the transport of angular
momentum and energy.

\end{abstract}
\keywords{ accretion -- hydrodynamics -- black hole physics -- outflows  -- 
galaxies: active -- methods: numerical} 

\section{Introduction}

Some of the most dramatic phenomena of astrophysics, such as quasars and 
powerful radio galaxies, are most likely powered by accretion onto supermassive
black holes (SMBHs). Nevertheless, SMBHs appear to spend most of their time 
in  a remarkably
quiescent state. SMBHs are embedded in the relatively dense environments of
galactic nuclei, 
and  it is natural to suppose that the gravity due to an SMBH 
will draw in matter  at high rates, leading to a high system luminosity.
However, this simple prediction often fails as many systems
are much dimmer than one would expect.

To illustrate a key problem in constructing theoretical models
for accretion onto a black hole, let us express
the luminosity due to accretion as  
\begin{equation}
L= \eta c^2 \MDOT_a,
\end{equation}
where we invoke
the simplest assumption, that the luminosity is proportional
to the mass accretion rate, $\MDOT_a$, and an efficiency factor, $\eta$.
The accretion luminosity is very uncertain because
$\eta$ is uncertain: 
$\eta$
ranges from $\sim 10^{-1}$
in a standard, radiatively efficient thin disk, 
to $\sim 10^{-11}$ for spherically symmetric accretion from 
a low density medium 
(e.g., Shakura \& Sunyaev 1973; Shapiro 1973; M\'{e}sz\'{a}ros 1975). 
The mass accretion rate  is also a source of uncertainty  in
estimates of the accretion luminosity because  $\MDOT_a$ depends on
the physical conditions and geometry at large distances from the black hole.
Nevertheless, it is customary to adopt
the analytic formula due to Bondi (1952)
to estimate the mass accretion rate. In his classic paper, Bondi (1952)
considered spherically symmetric accretion from 
a non-rotating polytropic gas with uniform density $\rho_\infty$
and sound speed $c_\infty$ at infinity. Under these assumptions, 
a steady state solution to the equations 
of mass and momentum conservation exists 
with a mass accretion rate of
\begin{equation}
\MDOT_B= \lambda 4 \pi R^2_B \rho_\infty c_\infty,
\end{equation} 
where $\lambda$ is a dimensionless parameter
that, for the Newtonian potential, depends only on the adiabatic index. 
The Bondi radius, $R_B$, is defined as
\begin{equation}
R_B=\frac{G M}{c^2_\infty},
\end{equation}
where $G$ is the gravitational constant and $M$ is the mass of the accretor.

The Bondi accretion formula predicts that SMBHs in typical galaxies should 
be more luminous than observations indicate when $\eta$ is assumed to
be as large as in a standard, radiatively efficient thin disk 
(e.g., Di Matteo et al. 1999, 2000, 2001; Loewenstein et al. 2001; 
Baganoff et al. 2001). In the context of equation (1),  one possible 
explanation for this disagreement is that the black hole accretion flow can be
radiatively inefficient because binding energy dissipated in the gas is
advected through the event horizon before being radiated
(Ichimaru 1977; Rees et al. 1982; Narayan \& Yi 1994, 1995; 
Abramowicz et al. 1995).
However, pure advection-dominated inflow may not be the whole story.
Even before recent observations forced us to 
confront very low SMBH luminosities, 
theorists had begun to realize that rotating, 
radiatively inefficient hydrodynamical (HD) flows are subject to strong 
convection (Begelman \& Meier 1982; Narayan \& Yi 1995), 
which can fundamentally change the flow pattern and its radiative properties 
(Blandford \& Begelman 1999; Quataert \& Narayan 1999; Narayan, Igumenshchev 
\& Abramowicz 2000; Quataert \& Gruzinov 2000). The theoretical studies showed 
that convection alters the steep ($\propto r^{-3/2}$) density profiles of 
the advection-dominated flow into a much flatter ($\propto r^{-1/2}$) profile, 
which can explain the faintness of many SMBHs 
because it predicts relatively low density close to the black hole (i.e.,
$\MDOT_a$ is low in eq.~2). Similar structural changes occur in 
the magnetohydrodynamical (MHD) limit (Stone \& Pringle 2001; 
Hawley, Balbus, \& Stone 2001; Machida, Matsumoto \& Mineshige 2001; 
Igumenshchev \& Narayan 
2002; Hawley \& Balbus 2002), although here the turbulence is probably
driven by  magnetorotational instability  (MRI) rather than thermal convection
(Balbus \& Hawley 2002; but see Abramowicz et al. 2002 and 
Narayan et al. 2002 for alternative views).

The turbulent character of both HD and MHD models does not
settle the issue of what happens to the energy and angular momentum
that must be transported away. There are two possibilities:
(i) turbulent transport effectively shuts off the accretion flow,
turning it into a closed circulation (Narayan et al. 2000; Quataert \&
Gruzinov 2000) or (ii) turbulent transport drives powerful outflows
that can strongly modify the black hole's environment 
(Narayan \& Yi 1994, 1995; Blandford \& Begelman 1999).
Recent MHD simulations bring new insights that may help us
to resolve this issue. For example, Hawley \& Balbus's (2002) 
three-dimensional MHD simulations show that, with and without resistive
heating, mass and energy in  nonradiative accretion flows are carried
off by an outflow in keeping with the outline of the second
possibility.

Another possible solution to the problem of very low SMBH luminosity  
is that mass is captured into the accretion flow at a rate that
is far lower that $\MDOT_B$. Thus, a complete formulation of the accretion 
flow must also consider the outer boundary conditions. SMBHs draw matter from 
an extended medium and most authors assume that the Bondi (1952) formula 
provides 
an adequate approximation for the rate of mass supply. There has  been little 
systematic work done to demonstrate that this assumption is justified and 
correct. Igumenshchev \& Narayan (2002) showed that even non-rotating Bondi 
accretion can be altered, in particular that the mass accretion rate can be 
reduced below the Bondi rate. The cause of the $\MDOT_a$ reduction in 
Igumenshchev \& Narayan's simulations is
the local generation of energy through  by magnetic reconnection,
which leads to the development of efficient convection.
However, in  Igumenshchev \& Narayan's simulations
the flow was supersonically injected into the computational grid
through the outer boundary, at  a specified rate (determined
from the Bondi formula).  
Their calculations did not address self-consistently the problem
of the mass accretion and supply rate because the latter was fixed
at the outer boundary.
As we mention above, the Bondi formula has been derived under 
the assumption that the gas is non-rotating and only under the influence of 
the central gravity. Thus, for a given gravitational field,
the gas internal energy determines the accretion rate. By relaxing
this assumption, introducing additional forces or sources of energy, 
one may find that the mass supply rate is much lower than the one predicted by
the Bondi formula. For example, 
the rate at which matter is captured by a black
hole can be severely limited when the matter is heated by X-rays produced
near the black hole (Ostriker et al. 1976) or by  mass outflow
from the central region (Di Matteo et al. 2002). 
In these two cases, the gas internal energy
is increased. Introducing kinetic energy to the gas at infinity
may have a similar effect:  
although the flow outside the Bondi accretion radius often can be described
as nonrotating, even a tiny amount of angular momentum, $l$ --- when followed
inward --- could severely limit the rate at which matter is captured by the
black hole. 

Our focus shall be on assessing the gross properties of rotating accretion 
flows onto  black holes. We consider a classic Bondi accretion flow with 
the only modifications being the introduction of a small, 
latitude-dependent angular momentum at infinity and
a pseudo-Newtonian gravitational potential. 
The imposed angular momentum is small enough  to have a negligible
effect on the density distribution at the outer boundary, which remains
spherically symmetric. Therefore, 
matter near the rotational axis can be accreted.
We thus consider a very simple model of an accretion flow, 
far simpler than those occurring 
in nature, as we neglect the gravitational field due to the host galaxy,
radiative heating and cooling, viscosity and MHD 
effects. For example, we will not consider here the transport of energy and 
angular momentum outward, as needed to accrete matter with a specific angular 
momentum higher than $2 R_S c$ 
(where $R_S=2 G M/c^2$ is the radius of a Schwarzschild black hole).
Nevertheless, the results presented here provide a useful exploratory
study of accretion onto black holes. In particular, our results constitute
a ``baseline'' for evaluating the effects of dissipative and transport
processes in subsequent work. 
They also serve as a ``proof-of-concept'' for the reduction
of the mass accretion rate due to a small angular momentum
in the accretion flow. 

\subsection{Expectations}

Before we embark on a detailed analysis and numerical HD 
simulations, we first consider the problem of  accretion of  low-$l$ 
material in a general way. If the matter far from the SMBH
(well outside the Bondi radius)
has a uniform density and a specific angular momentum, $l$, which
exceeds $2R_Sc$ --- a tiny value
compared to the Keplerian angular momentum at $R_B$ --- then
no accretion will take place without angular momentum transport.
After a transient episode of infall, the gas will pile up outside
the black hole and settle into a nearly steady state atmosphere
bounded by a centrifugal barrier near the rotation axis. Realistically,
matter far from the SMBH will have a range of angular momenta, and
in a steady state with axisymmetry, there will always be 
low-$l$ material close to the axis that can accrete
steadily through a funnel along the rotational axis.

For this highly idealized problem, one would expect that the mass accretion
rate should scale with the dependence of the angular momentum on the polar
angle, $\theta$, at the outer radius, $r_o$: 
the larger the amount of the material with $l > 2 R_S c$
at $r_o$, the lower the mass accretion rate. This geometrical argument
on the nature of  the $\MDOT_a$ $vs.$ $l$ relation can be quantified 
as follows. If the angular momentum  depends on $\theta$ as 
\begin{equation}
l(\theta)=l_0 f(\theta),
\end{equation}
where  $f=1$ on the equator ($\theta=90^\circ$) and monotonically
decreases to zero at the poles ($\theta=0^\circ$ and $180^\circ$), 
then a naive expectation would be that  $\MDOT_a/\MDOT_B$ scales with 
the solid angle within which $l< 2 R_S c$:
\begin{equation}
\frac{\MDOT_a}{\MDOT_B}=\frac{\Delta \Omega_o}{4\pi}= 1 - \cos{\theta_o},
\end{equation}
where $\theta_o$ is the width of the angular distribution for which 
$l \le 2 R_S c$. The latter can be formally defined as 
\begin{equation}
\theta_o \equiv f^{-1} \left[\min\left(1,\frac{2R_Sc}{l_0}\right)\right],
\end{equation}
where $f^{-1}$ represents the functional  inverse  of $f$. This simple 
geometrical argument is based on the assumption of radial flow and implies 
that if $l\leq 2 R_S c$ the material will be accreted approximately at 
the Bondi rate. If relation (6) were true then the accretion rate should 
decrease with increasing $l_0$ for a fixed $R_S$. Additionally, one would 
expect that $\MDOT_a/\MDOT_B$ should be independent of $R_B$ for fixed $R_S$ 
and $l_0$. However, to determine the actual mass accretion rate even in 
this idealized case we need to perform numerical simulations, 
as just HD effects of the inviscid fluid
make the above geometrical argument invalid both quantitatively
and qualitatively.

The  main result of our numerical HD calculations is that
the properties of the accretion flow do not depend as much on 
the outer boundary conditions (i.e., the amount as well as distribution
of the angular momentum) as on the geometry of the {\it non-accreting matter}. 
Material with $l \simgreat 2 R_S c$ cannot accrete and forms a thick torus 
near the equator. This thick torus and its formation have been a subject of 
numerous studies (see below).  Our simulations  show that the dependence of 
angular momentum on $\theta$ in the torus gets weaker with decreasing radius.
The material with $l \simgreat 2 R_S c$ inflows in the polar region,
turns around as it reaches a centrifugal barrier, and then starts to
outflow  along the equator. As a result, a thick torus of nearly
uniform specific angular momentum forms as gas of 
the highest angular momentum ($l \approx l_0 > 2 R_S c$) near the equator 
is replaced by gas of lower angular momentum ($l \simgreat 2 R_S c$). 
The geometry of the polar region, where material is accreted (the funnel)
and the mass accretion rate through it are constrained by the size and 
shape of the torus. We describe the size of the torus by an
angle, $\theta_t$,  
between the torus's upper envelope and the pole at a characteristic
radius.

Our HD models show that the $\MDOT_a$ $vs.$ $l$ relation
has three regimes for a given $f(\theta)$:
(i) for low $l$ (i.e., $l < 2 R_S c$ for all $\theta$ at large radii), 
the torus does not form and $\MDOT_a= \MDOT_B$,
(ii) for intermediate $l$, or more appropriately where there is 
a narrow range of $\theta$ for which $l>2 R_S c$ so $\theta_o > \theta_t$, 
$\MDOT_a \sim const$ with
the actual value of the  constant depending on the ratio $R_S/R_B$ 
and (iii) for  high $l$, or in the case for which 
$l> 2 R_S c$ at nearly all $\theta$ so $\theta_o < \theta_t$,  
$\MDOT_a$  decreases with increasing $l_0$.

Comparing the $\MDOT_a$ $vs.$ $l$ relation based on  our HD
models with that described by equation (5), we find that the two
relations agree exactly in the first regime, disagree qualitatively
and quantitatively in the second regime 
[the HD models predict $\MDOT_a$ lower than eq. (5)], 
and agree again but only
qualitatively in the third region 
[the HD models predict $\MDOT_a$ higher than eq. (5)].
Thus the geometrical argument used above does not hold.
However, we can use a modified version to describe 
the key aspects of  the $\MDOT_a$ $vs.$ $l$ relation.
The modification to the geometrical argument is to replace 
the solid angle within which $l< 2 R_S c$
at the outer boundary, $\Delta \Omega_o$, 
with the solid angle within which $l< 2 R_S c$
at {\it a characteristic radius comparable with the sonic radius},
$\Delta \Omega_f$ (i.e., the solid angle of the accretion funnel). 
Thus, $\MDOT_a/\MDOT_B \approx \Delta \Omega_f/ 4\pi$.  
The insensitivity of $\MDOT_a$  to the angular momentum at
infinity, in the second regime, can be attributed to 
the relative insensitivity of the torus, and thus the funnel, to the angular
momentum distribution at infinity. In terms of the solid angle within
which $l < 2 R_S c$, this corresponds to
$\Delta \Omega_f=constant$  for variable $l_0$, provided that
$\Delta \Omega_f < \Delta \Omega_o$ (i.e., $\theta_o > \theta_t$).  
On the other hand, 
the decrease of $\MDOT_a$ with increasing $l_0$, in the third regime, 
can be attributed to the fact
that  $\Delta \Omega_f$ decreases with increasing $l_0$, provided that
$\Delta \Omega_f > \Delta \Omega_o$ ($\theta_o < \theta_t$).  
We find that the mass accretion rate decreases with increasing $l_0$ 
more slowly  than predicted by eq. (5) 
in the third regime because the sonic point radius starts
to increase as the funnel gets narrower.

\subsection{Previous Work}

Similar calculations have been performed before. For example, the formation 
of rotationally supported thick tori from inviscid accretion of gas with 
various initial angular momentum distributions has been reported 
(Hawley, Smarr \& Wilson 1984a; 1984b; Clarke, Karpik \& Henriksen 1985; 
Hawley 1986; Molteni,  Lanzafame \& Chakrabarti 1994; 
Ryu et al. 1995; Chen et al.  1997).
However, there is one key  difference between our work and some past work: 
our outer radial boundary is located outside the Bondi radius and we adopt 
subsonic, Bondi-like outer radial conditions whereas  Molteni et al. 1994, 
Ryu et al. 1995, and Chen et al.  1997 (see also Toropin et al. 1999; 
Kryukov et al. 2000; and Igumenshchev \& Narayan 2002) imposed  outer 
boundary conditions inside the Bondi radius or even inside the sonic radius.
The latter approach allows one to study cases where $R_S/R_B$ is as low as 
in some real systems (e.g., $R_S/R_B=10^{-5}$ in Chen et al. 1997) but 
this approach is not suitable for addressing our main issue: what is the mass 
supply rate. The approach adopted by Hawley et al. (1984a, 1984b), 
Clarke et al. (1985) and Hawley (1986) is much closer to ours 
as they also used subsonic outer boundary conditions. However, these
authors did not consider how the accretion rate onto the black hole depends 
on the angular momentum distribution beyond the Bondi radius but rather 
focused on the formation of the thick torus. As far as the treatment of 
the outer radial boundary is concerned, our simulations are also similar 
for those of Ruffert (1994), who studied three-dimensional hydrodynamic 
Bondi-Hoyle accretion. 

Other studies are also relevant to our work. Several authors considered 
accretion onto black holes with a focus on the evolution of rotationally 
supported thick tori including the transport of angular momentum and energy 
(e.g., Igumenshchev \& Abramowicz 1999; Stone, Pringle \& Begelman 1999;
Stone \& Pringle 2001; Machida et al. 2001; 
Hawley \& Balbus 2002). The main difference between our work and these studies 
is that the other authors adopt the point of view that virtually all of 
the material at large radii has too much angular momentum to be accreted 
without the transport of angular momentum. They assume that material with 
zero or very low angular momentum is unimportant dynamically, and that 
accretion is dominated by  angular momentum and energy transport processes.
For example, for their initial conditions Stone et al. (1999) and  
Hawley \& Balbus (2002) adopted a bounded torus in hydrostatic equilibrium 
with constant angular momentum, embedded in zero angular momentum ambient gas
which is also in  hydrostatic equilibrium. Thus, these  calculations were set 
up so that, if not for the transport of angular momentum, there would be 
neither time evolution nor  mass accretion. We recognize that transport 
processes are essential, but assert that the zero or very low angular momentum 
material also deserves a rigorous treatment because it can play an important 
role in determining the total mass supply  and accretion rate.

In this paper, we consider a far simpler case of an accretion flow
than those occurring in nature (see Section~4). For example, we neglect 
viscosity and MHD effects. In particular, 
the MRI has been shown to be a very robust and universal mechanism to produce 
turbulence and the transport of angular momentum in disks at all radii 
(Balbus \& Hawley 1998).

The outline of this paper is as follows. We describe our calculations 
in Section 2. In Section 3, we present our results. We summarize our results 
and discuss them together with their limitations in Section 4.

\section{Method}

\subsection{Hydrodynamics}

To calculate the structure and evolution of an accreting flow, we solve 
the equations of hydrodynamics
\begin{equation}
   \frac{D\rho}{Dt} + \rho \nabla \cdot {\bf v} = 0,
\end{equation}
\begin{equation}
   \rho \frac{D{\bf v}}{Dt} = - \nabla P + \rho \nabla \Phi,
\end{equation}
\begin{equation}
   \rho \frac{D}{Dt}\left(\frac{e}{ \rho}\right) = -P \nabla \cdot {\bf v},
\end{equation}
where $\rho$ is the mass density, $P$ is the gas pressure, 
${\bf v}$ is the velocity, and $e$ is the internal energy density.
We adopt an adiabatic equation of state
$P~=~(\gamma-1)e$, and consider models with $\gamma=5/3$.
Our calculations are performed in spherical polar coordinates
$(r,\theta,\phi)$. We assume axial symmetry about the rotational axis
of the accretion flow ($\theta=0^\circ$). 

We present simulations  using 
the pseudo-Newtonian potential $\Phi$ 
introduced by Paczy\'{n}ski \& Wiita (1980) 
\begin{equation}
\Phi=-\frac{G M}{r-R_S}.
\end{equation}
This potential approximates general relativistic effects
in the inner regions, for a nonrotating black hole. 
In particular, the Paczy\'{n}ski--Wiita potential
reproduces the last stable circular orbit at $r=3 R_S$
as well as  the marginally bound orbit at $r=2 R_S$.

\subsection{Initial conditions and boundary conditions}

For the initial conditions we adopt a Bondi accretion flow with zero
angular momentum everywhere except for the outermost part of the flow.
In what follows we briefly review the basics of Bondi accretion
that  allow us to specify details of our initial and boundary conditions
as well as to interpret our results.

The Bernoulli function can be written as
\begin{equation}
{B}= H+ \frac{v_r^2+v_\theta^2}{2} +\frac{l^2}{2 r^2 \sin^2{\theta}}
+ \Phi,
\end{equation}
where H is the enthalpy and $l=v_\phi r \sin{\theta}$ 
is the specific angular momentum.
For a polytropic equation of state $P=K \rho^\gamma$,
the polytropic constant can been expressed as 
$K=\rho^{1-\gamma}_\infty c^2_\infty/\gamma$,
where $c_\infty$ is the sound speed at infinity [i.e., $c^2_\infty\equiv 
(dP/d \rho)_\infty$]. Therefore the enthalpy becomes
\begin{equation}
H=\int_{\rho_\infty}^\rho \frac{dP}{\rho}=
\frac{ \gamma}{\gamma-1}\left(\frac{P}{\rho}-\frac{P_\infty}{\rho_\infty}\right)=
\frac{ 1}{\gamma-1}c_\infty^2\left[\left(\frac{\rho}{\rho_\infty}\right)^{\gamma-1}-1\right].
\end{equation}

Let us consider spherically symmetric, steady-state  
Bondi accretion onto an object with a Paczy\'{n}ski--Wiita (PW, hereafter) 
potential.
In such a case, $v_\theta=l=0$ and
the Bernoulli function simplifies to
\begin{equation}
B=-\frac{GM}{r-R_S}+\frac{c_\infty^2}{\gamma-1}\left[\left(\frac{\rho}
{\rho_\infty}\right)^{\gamma-1}-1\right]+\frac{v_r^2}{2}=0.
\end{equation}

Introducing dimensionless variables
$\alpha=\rho/\rho_\infty$, $x=r/R_B$, $R'_S=R_S/R_B$, 
and $u_r=v_r/c_\infty$, we rewrite eq. (13) as
\begin{equation}
B'=-\frac{1}{x-R'_S}+\frac{1}{\gamma-1}(\alpha^{\gamma-1}-1)+\frac{u_r^2}{2}=0.
\end{equation}
The transonic solution of the Bondi problem yields
the sonic point at
\begin{equation}
x_s=\frac{5-3\gamma+8 R'_S+\sqrt{\Delta}}{8},
\end{equation}
where
\begin{equation}
\Delta=(5-3\gamma+8 R'_S)^2-64 (1-\gamma+R'_S) R'_S. 
\end{equation}
The dimensionless
mass accretion rate $\lambda\equiv x^2 \alpha u_r$ for the transonic
solution can be calculated as
\begin{equation}
\lambda=x_s^2 \alpha_s u_{r,s}=x_s^2 \alpha_s^{(\gamma+1)/2},
\end{equation}
with the dimensionless density at the sonic point, $\alpha_s$, given by
\begin{equation}
\alpha_s=\left[\frac{1}{2}+\frac{2(x_s-R'_S)}{x_s}-\frac{\gamma}{2}\left(4\frac{x_s-R'_S}{x_s}-1\right)\right]^{-1/(\gamma-1)}.
\end{equation}
For $\gamma=5/3$, 
the dimensionless sonic radius and mass accretion rate are
\begin{equation}
x_s=R'_S+\sqrt{2 R'_S/3}
\end{equation}
and
\begin{equation}
\lambda=\left(\frac{3}{4} R'_S\right)^2\left(1+\sqrt{\frac{2}{3 R'_S}}\right)^4,
\end{equation}
respectively.
Note that, for the PW potential as well as the fully general
relativistic problem (Begelman 1978), Bondi flow with $\gamma=5/3$ has
a sonic point at roughly the geometrical mean between the Bondi
radius and the Schwarzschild radius. This is in contrast to the purely
Newtonian case in which $x_s=(5-3\gamma)/{4} \rightarrow 0$
as $\gamma  \rightarrow 5/3$.

We will use the results for Bondi accretion with the PW
potential [i.e., equations (17)-(20)] as a reference point for analyzing
our results and to set the initial conditions. Namely,
we adopt $v_\theta=0$ while $v_r$ and $\rho$
are computed using the Bernoulli function and mass accretion rate
$\MDOT_B\equiv - 4 \pi r^2 \rho v_r=  
\lambda 4 \pi R_B^2 \rho_\infty c_\infty$.
We set $\rho_\infty=1$ and specify $c_\infty$ through
$R'_S$ (note that $R'_S=2c^2_\infty/c^2)$.
We complete specification of the initial conditions by adopting a non-zero
specific angular momentum $l$ for $r \ge x_s R_B$.
We  also ran a few models in which the
specific angular momentum initially 
is non-zero only at the outer radial boundary $r_o$. 
We found that, except for the initial transient, models with these
initial conditions for $l$ give the same results. However,
adopting a non-zero specific angular momentum $l$ for $r \ge x_s R_B$
reduces 
the computational time significantly  because the initial
transient lasts for a much shorter time.

We consider a general case where the angular momentum depends
on the polar angle  via
\begin{equation}
l(r_o,\theta)=l_0 f(\theta),
\end{equation}
with $f=1$ on the equator ($\theta=90^\circ$) and $f=0$
at the poles ($\theta=0^\circ$ and $180^\circ$). We express the angular momentum
on the equator  as
\begin{equation}
l_0=\sqrt{R'_C} R_B c_\infty,
\end{equation}
where $R'_C$ is the ``circularization radius'' on the equator in units of $R_B$
for the Newtonian potential (i.e.,  $ GM/r^2= v^2_\phi/r$ at $r= R'_C R_B$).

We adopt three forms for the function $f(\theta)$: 
\begin{equation}
f_1(\theta)=1-|\cos\theta|,
\end{equation}
\begin{equation}
f_2(\theta)=1-\cos^{10}\theta, 
\end{equation}
and
\begin{equation}
f_3(\theta)=\left\{ \begin{array}{ll}
                 0    & {\rm
                  for}~~\,~~\theta~~<~~\theta_o~~{\rm and}~~\theta >
                 180^\circ -\theta_o  ~~\,~~\\
                 l_0  & {\rm for}~~\,~~\theta_o\le~~\theta~~\le 
                   180^\circ-\theta_o.
\end{array}
\right.
\end{equation}
Our choice of the $\theta$ dependence of $l$ at infinity is motivated
by the following:
the $l$ distribution described by eq. (23) allows us to study
a case in which  the flow at large radii is very similar to the Bondi
flow as the material with $l> 2 R_S c$ is confined to within a narrow
range of $\theta$ above and below the equator if $l_0$ is 
close to $2 R_S c$. In such a case
most of the material at large radii has $l< 2 R_S c$ and can be accreted,
in principle. On the other hand, the $l$ distribution described 
by eqs. (24) or (25) allows us to study the opposite case, in which  
the material with $l> 2 R_S c$  at large radii 
occupies nearly the entire range of 
$\theta$ above and below the equator, even if $l_0$ is not much larger than
$2 R_S c$. Then  
most of the material at large radii has $l> 2 R_S c$;
only in a narrow polar region is $l$ low enough for accretion to take place
($l< 2 R_S c$). We note that for 
the $\theta$ dependence of $l$ described by $f_1$ and $f_2$, 
we calculate $\theta_0$ using eq. (6) and an assumed value for 
$l_0$. For the $\theta$ dependence of $l$ described by $f_3$,
$l_0$ as well as $\theta_0$ are free parameters.

Our standard computational domain is defined to occupy the radial range
$r_i~=~1.5~R_S \leq r \leq \ r_o~=~ 1.2~R_B$ and
the angular range
$0^\circ \leq \theta \leq 180^\circ$. We consider models with $R'_S$ 
from $10^{-2}$ to $10^{-3.5}$.
The $r-\theta$ domain is discretized into zones.  
For $R'_S=10^{-2}$, $10^{-3}$, and $10^{-3.5}$,  
our  numerical resolution consists of 100, 140, and 160 zones in 
the $r$ direction, respectively.
In the $\theta$ direction, our numerical resolution consists
of 100 zones for all values of $R'_S$. We usually fix zone size
ratios, $dr_{k+1}/dr_{k}=1.05$, 
$d\theta_{l}/d\theta_{l+1} =1.05$ for $0^\circ \le \theta \le 90^\circ$ 
(i.e., the zone spacing is decreasing in this region)
and $d\theta_{l+1}/d\theta_{l} =1.05$ for $90^\circ \le \theta \le 180^\circ$.
For runs with very high $l_0$ we adopt $d\theta_{l}/d\theta_{l+1} =1.0$ for
$0^\circ \le \theta \le 180^\circ$ (see section 3.3).
For runs with a very narrow funnel in which we use
a step function for $f(\theta)$, we adopt
$d\theta_{l+1}/d\theta_{l} =1.03$ for $0^\circ \le \theta \le 90^\circ$ 
(i.e., the zone spacing is increasing in this region)
and $d\theta_{l}/d\theta_{l+1} =1.03$ 
for $90^\circ \le \theta \le 180^\circ$.

The boundary conditions are specified as follows. At the poles,
(i.e., $\theta=0^\circ$ and $180^\circ$), we apply an axis-of-symmetry 
boundary 
condition.  For the inner and outer radial boundaries, we apply an outflow 
boundary condition.  Our choice for the location of the inner radial
boundary, $r_i=1.5 R_S$, ensures that the flow near this boundary
is supersonic and the outflow condition is appropriate.
To represent steady conditions at the outer radial
boundary, during the evolution of each model we continue to apply the
constraints that in the last zone in the radial direction, $v_\theta=0$,
$v_\phi=l_0 f(\theta)/ r \sin{\theta}$, and the density is fixed at the 
Bondi value at all times. Note that we allow $v_r$ to float.
We have found that this technique, when applied
to calculations of spherically symmetric accretion, produces
a solution that relaxes to the steady-state Bondi solution for $l=0$.
For  non-zero $l$, we find that
the outer radial boundary must be sufficiently far 
from the circularization radius to ensure
that the assumption of pure radial flow (i.e.,
$v_\theta(r_o)=0$) does not change the solution (see Section 3). 

To solve eqs. (7)-(9) we use the ZEUS-2D code described by Stone \& Norman 
(1992), modified  to implement the PW potential. As with all Eulerian
codes, one ought to test how much numerical diffusion error 
(``numerical viscosity'') affects the results. In particular, shocks are 
captured with a standard quadratic artificial viscosity 
(see Stone \& Norman 1992). Since we compute the evolution
of adiabatic flows without any physical viscosity,
numerical effects can, in principal, limit the accuracy
of the results.  The artificial viscosity, which is not shown
in our equations (7)-(9), is present only when the (inward) velocity
divergence is nonvanishing. When present, artificial viscosity heats 
the gas and transports angular momentum. 
As a test, we reran  model A04fl1a with a resolution of 150 in the $r$
direction and of 200
in the $\theta$ direction (i.e., the resolution of the test run is twice that
in model A04fl1a in both directions).
We found that the mass accretion rate in model A04fl1a and its
higher resolution counterpart agree to within one part in $10^3$.
As for heating of the gas due to artificial viscosity, we found that 
the polytropic constant, $K$, is conserved in all of our simulations, 
except for departure of $\simless 1\%$  near the equator for  
small radii, where weak shocks appear. We conclude
that our numerical simulations treat inviscid flows satisfactorily.

\section{Results}

Our numerical models are specified by several parameters.
The length scale is determined by the black
hole radius in units of the Bondi radius, $R'_S$. 
Our second parameter is the adiabatic index, $\gamma$.
The last parameter (or a function rather) is the angular
momentum at the outer radial boundary, $l=l_0 f(\theta)$.

In real systems, $R'_S$ is  relatively small 
($10^{-8} \simless R'_S=2c^2_\infty/c^2 \simless 10^{-5}$) because
the sound speed at large distances from a black hole
is very small compared to the speed of light. Models with so great
a radial domain would be very demanding computationally. 
We therefore decided to explore the main features of our model
by considering a smaller radial domain, $10^{-3.5}\leq R'_S\leq10^{-2}$.
We consider adiabatic flows with $\gamma=5/3$.
Finally, we assume an angular momentum distribution at the outer radial
boundary  as  described in Section~2. We focus our attention on accretion 
of matter with low angular momentum, i.e., 
where the corresponding centrifugal force  is small compared
to  gravity  for all $\theta$ at the Bondi radius:
\begin{equation}
\frac{l^2_0 f^2(\theta)}{R^3_B \sin^2{\theta}}< \frac{GM}{(R_B-R_S)^2}.
\end{equation}
In the limit  $R_B \gg R_S$ and using our definition of $R'_C$, we can
rewrite the above equation as
\begin{equation}
\frac{f^2(\theta)}{\sin^2{\theta}} < \frac{1}{R'_C}.
\end{equation}

Table~1 summarizes the properties of the simulations we discuss
here. Columns (2) through (7) give the numerical resolution in 
the radial direction;
the black hole radius compared to the Bondi radius, $R'_S$;
the circularization radius compared to the Bondi radius, $R'_C$;  
the specific angular momentum on the equator at $r=r_o$, $l_0$, in
units of $2 R_S c$; 
the width of the angular distribution for which $l \le 2 R_S c$, $\theta_o$;
and the angular momentum dependence on the polar angle at the outer
boundary, $f(\theta)$,  respectively.
Table~1 also presents the final time at which we stopped each simulation
(all times here are in units of the Keplerian orbital time at $r=R_B$),
the maximum specific angular momentum at the inner radial boundary, 
$l_{a}^{max}$, and 
the time-averaged mass accretion rate through the inner radial boundary
measured near the end of the simulation,  in units of  the corresponding 
Bondi accretion rate. Finally, column (11) gives comments about runs different
from the standard runs (e.g., the outer radial boundary
set at $12 R_B$ instead of $1.2 R_B$)

Our simulations show that for $l_0> 2R_S c$ 
the accretion flow consists of a thick, equatorial torus where the material
has too much angular momentum to be accreted, and a supersonic polar funnel 
where the material has $l$ low enough to be accreted.
We describe an example of such an accretion flow in some detail first
(Section 3.1). This is followed by a limited parameter survey in
which we focus on varying three key aspects of our models:
the maximum angular momentum for a given
distribution, the angular distribution of angular momentum on 
the outer boundary, and the black hole radius compared to the Bondi radius.

\subsection{Accretion flow consisting of a torus and a funnel}

In this section we describe the properties
and behavior of our model in which $R'_S=10^{-3}$, $R'_C=10^{-1}$, and
$f(\theta)=1-|\cos{\theta}|$ (model B04f1a). For 
the above parameters, the specific angular momentum on the equator at 
the outer boundary is $\sim 3.5$ in units of $2 R_S c$, thus the material 
that cannot be accreted onto the black hole is located relatively
close to the equator,  $45^\circ \le \theta \le 135^\circ$ 
at the outer boundary.

Figure~1 presents a sequence of  specific angular
momentum contours  and  velocity fields for model B04f1a.
The length of the arrows in the lower panels is proportional to
the poloidal velocity $\sqrt{v^2_r+v^2_\theta}$.
As we mentioned in Section~1,
after a transient episode of infall, the gas with $l> 2 R_S c$  
piles up outside the black hole and settles into a thick torus
bounded by a centrifugal barrier near the rotation axis. 
The low-$l$ material close to the axis accretes almost
steadily through a funnel. The distribution of  specific angular momentum 
in the torus becomes increasingly homogeneous as  material
with $l\sim 0.9$ replaces material with higher $l$.
The flow in the torus is subsonic, variable and is directed outward near
the equator and inward close to the poles. 
There is no symmetry with respect to the equator. The time-dependence
persists even after $t_f$=9. Thus, a meridional circulation is established
in the torus.
However, it is important to note
that the time-averaged gross properties of the flow 
(such as the mass accretion rate
and the shape of the torus and funnel) settle down to a steady state.

To show the accretion flow in more detail, Figure~2 presents
the enlargement of two  panels from Figure~1 (the panels from the 
the end of our simulations). This figure  shows also 
the radial sonic surface, that is, the location where the radial Mach number
($M_r\equiv v_r/c_s$) equals one. Note that the specific angular momentum 
in the torus for small
radii is nearly constant ($l \sim 1$ in units of $2 R_S c$).
Additionally, for small radii the sonic surface approximately
coincides with the $l\approx 0.9\times 2 R_S c$ surface. The `south'
lobe of the sonic surface is more elongated than the `north' lobe
at the end of the simulations. However, in a time-averaged sense
both lobes are very similar. 

Next we consider the angular dependence 
of the flow at small radii. Figure~3 is a plot of the angular
dependence of the Mach number, specific angular momentum 
and mass flux density at $r=r_i=1.5 R_S$.
This figure shows three important features of the accretion flow:
(i) the accretion flow is highly supersonic in the radial direction, 
(ii) $l_a^{max}$ is not $2 R_S c$ but rather somewhat smaller
($0.9 \times 2 R_S c$) and (iii) the mass flux density is a strong function
of the polar angle: it is nearly flat near the poles and peaks
near the equator where $l\approx 0.6$ in units of $2 R_S c$. 

To provide more insight into the character of the accretion flow,
Figure~4 shows the angular dependence of flow  properties 
on the sonic surface (the $M_r=1$ surface).
The top panel shows the radius at which $M_r=1$, $r_s$.
The second panel from the top shows the specific angular momentum
and the latitudinal  Mach number, $M_\theta\equiv v_\theta/c_s$.
The second panel from the bottom  and  the bottom panel show, respectively, 
the mass flux density ($\rho v_r$) and a measure of how much the flow
deviates from a purely radial flow at the sonic surface, $\Delta \theta$.
We define $\Delta \theta$ as the difference between a given polar angle, 
$\theta$, on the sonic surface, and the polar angle at which the gas with 
a given $l$ originated at $r=r_o$. Because our models conserve specific 
angular momentum along streamlines, 
$\Delta \theta$ can be formally estimated as 
\begin{equation}
\Delta \theta(\theta)=\theta- f^{-1}(l(r_s,\theta)/l_0),
\end{equation}
for $0^\circ \le \theta \le 90^\circ$ and 
\begin{equation}
\Delta \theta(\theta)=180^\circ -\theta- f^{-1}(l(r_s,\theta)/l_0),
\end{equation}
for $90^\circ \le \theta \le 180^\circ$.

Figure~4 shows that the sonic radius is the highest along the poles, comparable
to the Bondi sonic radius [see eq. ~(19)] and decreases by approximately 
an order of magnitude near the equator (see also Figure~2). 
Additionally, Figure~4 shows that the flow deviates from  radial flow on 
the sonic surface. The deviation from  radial flow increases with
increasing angle from the poles. For angles from the poles 
$\simless 30^\circ$, the deviation  is small and the radial approximation  
holds.  However, the mass flux density is dominated by matter originating
at  angles from   $> 40^\circ$ from the poles, for which 
the deviation from radial flow can be large 
(i.e., $10^\circ < \Delta \theta < 50^\circ$).
We note that the total mass accretion rate is dominated by 
streamlines which deviate only moderately from radial flow on the sonic 
surface. This conclusion is based on the observation that 
$\Delta \theta \simless 10^\circ$ for  material
with $l \sim 0.6 \times 2 R_S c$.

Finally, to provide some insight into the time dependence, Figure~5 shows
the time evolution of the mass accretion rate in units of the corresponding
Bondi rate. Initially, $\MDOT_a$ drops from 1 to 0.23 at t~=~0.25.
Then it starts oscillating around 0.3 with an amplitude of 
$\sim 0.1~\MDOT_a/\MDOT_B$.

\subsection{Dependence of accretion flow properties on $l_0$}

To check the trends of the accretion flow with the main
model parameters, we have performed a limited parameter survey.
We first describe our results 
for models with all parameters the same as in model B04f1a
except for  $l_0$.

The main result from our runs with various $l_0$ is that the 
mass accretion rate does not depend  on this parameter (see Table~1), within
the constraints imposed by assuming 
low angular momentum at $R_B$ [eq. (27)].
The basic reason for this result is the fact that the properties of 
the accretion flow are determined by the geometry of the  torus,
which has a uniform specific angular momentum slightly larger than the
maximum $l$ of the accreted gas.
The material with $l \simgreat 2 R_S c$ inflows in the polar region,
turns around as it reaches
a centrifugal barrier, and then starts to outflow along the equator. 
As a result, a thick torus forms with  gas of the highest angular momentum 
($l \approx l_0 > 2 R_S c$) near the equator 
being replaced by gas of  lower angular momentum ($l \simgreat 2 R_S c$). 
The geometry of the polar region, where material is accreted (the funnel)
and the mass accretion rate through it are constrained by the shape of the
torus. 

To illustrate the insensitivity of the inner flow to the angular momentum
at large radii, Figure~6 shows the angular momentum contours for our model
B01f1a at $t_f=3.5$. For this model we assumed $l_0=2 R_S c$ so that in 
the purely radial case all material should be accreted. However, the flow
converges toward the equator, and  gas pressure effects allow only
the material with $l_0 \simless 0.87\times2 R_S c$ to be accreted.
The  material with  higher $l$ turns around and forms a thick torus with
properties very similar to those of model B04f1a, which has a 
much higher $l$ at large radii. Comparing  the left-hand 
panel of Figure~6 with the top panels of Figure~1 shows clearly
how different the angular momentum distributions are at large radii in 
these two runs. 
However, comparing  the right-hand  panel of Figure~6 with the
left-hand panel of Figure~2 shows that the angular
momentum distribution at small radii, 
the shape of the torus (the $l=0.9$ contour) as well
as the sonic surface are qualitatively similar for both runs.

We conclude that the relative insensitivity of the torus to the angular
momentum distribution at infinity is responsible for 
the fact that $\MDOT_a$ does not depend  on $l_0$ 
and that the geometrical argument we used in Section~1 is invalid.
As runs B04f1a and B01f1a illustrate,
the width of the $l$ distribution for $l < 2 R_S c$ at infinity 
does not control the mass accretion rate. Even if material with
low $l$ comes from close to the equator it still has to go `around'
the inner torus. We suggest that a more useful 
way of geometrically determining the mass accretion rate
is to compare  the angle between the torus upper envelope
and the pole, $\theta_t$,  and 
the width of the angular distribution for which $l \sim 2 R_S c$, $\theta_o$. 
However, such a method of determining $\MDOT_a$ poses the difficulty
that $\theta_t$ depends on radius and we 
need to know at what radius we have to measure $\theta_t$.

\subsection{Dependence of accretion flow properties on the angular
distribution of $l$}

Motivated by the conclusion from the previous section, we have
performed a few simulations using an angular distribution for $l$
that yields a smaller value of $\theta_o$ than that
given by $f_1$ (see eq. 23), 
for a given $l_0$. We wish to check whether
the insensitivity of $\MDOT_a$ to $l_0$, as described above,
holds in cases where $\theta_o$ is small, in particular when 
$\theta_o< \theta_t$. 

Inspection of the expression (27) shows that
there are three possible ways of decreasing
$\theta_o$ within the constraints imposed by the assumption of low angular 
momentum:
(i) reduce $R'_C$ by increasing $R_B$ for given $R_C$ and $R_S$
and increase $l_0$ at the same time,
(ii) choose a function $f(\theta)$ that 
increases with increasing $\theta$ very quickly for small $\theta$
and very slowly for larger $\theta$,
or  (iii) both.
We will be able to explore the first possibility only in a limited way 
(see next section) 
because, as we mentioned above, models with $R_B$  very high compared to 
$R_S$ are very demanding computationally. However, we can explore
the second possibility at relatively low computational cost for 
moderate $R_B$. 

As before, we consider in detail only one example: 
for our model B08f2a, all parameters are the same as for model B04f1a
with the exceptions that $l$ scales with $\theta$ at $r_o$ as
$1-\cos^{10}{\theta}$ and $d\theta_l/d\theta_{l+1}=1$. 
Such a flat distribution of $l$ for high $\theta$
reduces the size of the polar region 
where $l < 2 R_S c$ (i.e., the $\theta_o$ angle).
Thus, we expect that $\theta_o < \theta_t$ in this case and
the mass accretion rate will be lower than in
model B04f1a. Indeed, we find that $\MDOT_a$ in model B08f2a
is lower than that for model B04f1a by a factor of $\sim 3$.
We observe a similar reduction of $\MDOT_a$ for all runs using
$f_2$ compared to runs using $f_1$, regardless of $R_S/R_B$.

Our experiments with various $l_0$ and angular distributions 
of $l$ show that $\theta_o$ appears to be the key parameter
that determines the effect of the angular momentum at the outer boundary 
on the mass accretion rate through the inner boundary.
To illustrate the role of $\theta_o$, Figure~7 plots the mass accretion
rate as a function of ${\theta_o}$ for all our runs with
$R_S/R_B=10^{-3}$, without making any distinction
between models with different $f(\theta)$ and/or $l_0$.
Figure~7 shows one of the key results of this paper.
For large $\theta_o$ --- corresponding to a broad funnel 
at $R_B$ --- $\MDOT_a$ is constant, 
whereas for small  $\theta_o$ --- corresponding to a narrow funnel
at $R_B$ --- $\MDOT_a$ decreases with decreasing $\theta_o$.

Figure~7 also compares our numerical results (the solid line) with the simple
prediction, eq. (5) (the dashed line). Equation (5) predicts that $\MDOT_a$
decreases monotonically with decreasing $\theta_o$ for all $\theta$.
Generally, our prediction from Section~1 overestimates $\MDOT_a$
for large $\theta_o$ and underestimates $\MDOT_a$ for small $\theta_o$.
The decrease of $\MDOT_a$ with decreasing $\theta_o$ shown in our numerical
simulations is consistent with the theoretical prediction, but only
qualitatively and only  in the limit of a narrow funnel.

We note that in the limit of a narrow funnel the sonic radius increases 
with decreasing $\theta_o$.
We can understand the $\MDOT_a$ $vs.$ $\theta_o$ relation
in the limit of a narrow funnel by combining this increase in the sonic 
radius with two other facts: 
(i) the solid angle subtended by the accretion funnel 
decreases with increasing radius and 
(ii) the solid angle subtended by the accretion funnel
is greater than $\Delta \Omega_o$ ($\Delta \Omega_f=\Delta \Omega_o$ at $r_o$
by definition).
Thus,  $\Delta \Omega_f$ and consequently  $\MDOT_a$ decrease
with decreasing $\theta_o$. This explains why the decrease of $\MDOT_a$ with 
decreasing $\theta_o$ shown in our numerical
simulations is qualitatively consistent with the theoretical prediction.
The explanation for the quantitative difference is simply related
to the second fact listed above, i.e., 
$\Delta \Omega_f \ge \Delta \Omega_o$ at the sonic radius, $r_s < r_o$.

\subsection{Dependence of accretion flow properties on $R_S/R_B$ and $r_o$}

The most difficult parameters of our models to explore are 
the $R_S/R_B$ ratio and the outer radius of the computational domain.
As  mentioned earlier, we have to consider values of $R_S/R_B$ orders
magnitude larger than those found in astrophysical objects.
Similarly, exploring a wide range of $r_o$ is computationally prohibitive.
Therefore we have
performed only a few simulations in which we vary these parameters. 
We check  then whether our results  allow us to
extrapolate an $\MDOT$ $vs.$ $R_S/R_B$ relation 
to very low values of $R_S/R_B$.

Our results from the runs with various $r_o$
allow us to make a few observations. First, a proper study of Bondi accretion
with low angular momentum requires $R_S/R_B < 10^{-2}$.
For  $R_S/R_B = 10^{-2}$ or larger, 
the assumption of low angular momentum (eq. 27) is valid only
for a very narrow range of $l_0$ while the assumption
of radial flow at the outer boundary is valid only for $r_o$ much larger 
than the Bondi radius. For example, our models A03f1a and A03f1b
show that the location of the outer boundary can change
the solution dramatically for  $R_S/R_B < 10^{-2}$ and moderate
$l_0$ of $2.5\times 2 R_S c$. For these parameters the circularization
radius is relatively high, $R'_C=0.5$, and the assumption
that the flow is radial close to the Bondi radius is not valid.
Our models A03f1a and A03f1b differ only in the location of the
outer radial boundary, $r_o=1.2 R_B $ and $12~R_B$, respectively.
However, this difference produces dramatic change in the accretion
flow. For model A03f1b, the flow is non-radial even as far as the
Bondi radius and material with  $l$ as low as $\sim 0.4 \times 2 R_S c$
turns around near the equator and never makes it
through the inner boundary. Since the point where this material turns
is relatively far from the inner boundary, the sonic 
surface of the flow coming from the polar region
is approximately spherical and  almost  as large as the sonic surface in 
Bondi accretion with  $l=0$.
{\it Because of these properties of the sonic surface the mass
accretion rate is almost as high as the  Bondi rate}! 
This model once again demonstrates that the mass accretion
rate is determined more by the shape of the flow that does 
not accrete than by the angular distribution of $l$ at large
radii.
On the other hand, model A03f1a has $r_o=1.2$ and forces
the flow to be radial near the Bondi radius via our outer boundary
conditions. Therefore, the flow converges toward the equator
at smaller radii and the material that  goes through the inner boundary
has $l$ higher than in the model A03f1b. Additionally, the point
where the material with too high $l$ turns around near the equator
is closer to the black hole than the sonic point in the polar
region. Consequently, for model A03f1a the sonic surface
has the `figure-8' shape and the mass accretion rate is lower
than the Bondi rate.

Because of the sensitivity of the results to the location
of the outer boundary for models with $R_S/R_B = 10^{-2}$,
we do not consider these models in great detail.
However, we note that these models (with fixed $r_o$)
show an $\MDOT_a$ $vs.~l$ relation
very similar to the one we found for models with $R_S/R_B = 10^{-3}$.

We base most of our observations and conclusions on models with
$R_S/R_B = 10^{-3}$ not only because they are computationally
affordable but also because they do not depend on the location
of the outer boundary.  Comparing models B04f1a and B04f1b we
find that moving $r_o$ from $1.2~R_B$ to $12~R_B$ does
not change the solution for the accretion flow.
The reason for this is simply that, contrary to models with
$R_S/R_B = 10^{-2}$, the Bondi radius is large enough compared to $R_C$
that the $\theta$ component of  centrifugal force is relatively small. 
Consequently, the radial
approximation at the Bondi radius is  valid for 
models with $R_S/R_B = 10^{-3}$. 

Unfortunately, models with the ratio $R_S/R_B$ lower than
$10^{-3}$   are computationally too expensive to
be explored as fully as models with  $R_S/R_B=10^{-3}$.
For example, it takes about three weeks of CPU time on a modern workstation
to run one model with $R_S/R_B=10^{-3.5}$ for 3 dynamical
time scales at the Bondi radius. Therefore, we limited 
ourselves to two models with $10^{-3.5}$, model C04f1a and C08f2a.
Our primary goal of running these models was to check whether the mass
accretion rate would change compared to models with 
$R_S/R_B = 10^{-3}$.

Model C04f1a is meant to represent a model
with a wide accretion funnel (the second regime in the $\MDOT_a$ $vs.~l$
relation)
while model C08f2a is meant to represent a model with a narrow
accretion funnel (the third regime in the $\MDOT_a$ $vs.~l$ relation).
For model C04f1a, the mass accretion rate is smaller than
the corresponding model with $R_S/R_B=10^{-3}$ 
(i.e., model B04f1a, see Table~1)
but we note that the mass accretion rate was increasing with
time at the end of our calculations. Judging from the trend of
$\MDOT_a$ with time we conclude that the mass accretion rate
does not change much when the ratio $R_S/R_B$ decreases
from $10^{-3}$ to $10^{-3.5}$ in the wide funnel case.
We reach the same conclusion for the models in the narrow
funnel case. We recognize that our conclusions
are based on very limited data but they are consistent, at least
for small $R_S/R_B$,  with our
understanding of how the mass accretion rate is determined.

\section{Discussion}

This paper presents the first phase of our study of slightly rotating 
accretion flows onto  black holes. We decided to consider a far simpler case 
of an accretion flow than those occurring in nature. In particular, 
we neglected  the gravitational field due to the host galaxy, radiative 
heating and cooling, viscosity and MHD effects. Perhaps the most important 
simplification we made is neglecting the transport of energy and angular 
momentum outward as needed to accrete matter with a specific angular momentum 
higher than $2 R_S c$. Nevertheless, our results  provide a useful exploratory
study of accretion onto black holes as they have revealed unexpected 
properties and complexity of accretion flows in even this simple case.
Clearly, a lot more  work is needed to give a definitive answer to
the question of whether slow rotation of gas at large radii is really enough 
to reduce the mass accretion rate to the level required by observations. 
In what follows we will summarize our results,  briefly  review 
the limitations of our work, and discuss how the physical effects 
neglected here may change the results.

We have performed numerical 2D, axisymmetric, hydrodynamical simulations
of slightly rotating, inviscid accretion flows onto a black hole. We attempt 
to mimic the boundary conditions of classic Bondi accretion flows with
the only modifications being the introduction of a small, latitude-dependent 
angular momentum at the outer boundary and a pseudo-Newtonian gravitational 
potential. The distribution of $l$ with latitude allows the density 
distribution at infinity to approach spherical symmetry. The main result 
of our calculations is that the properties of the accretion flow do not depend 
as much on the outer boundary conditions (i.e., the amount as well as 
distribution of the angular momentum) as on the geometry of 
the {\it non-accreting matter}. Additionally, we find that the mass accretion 
rate $vs.$ angular momentum relation for a given angular distribution of $l$ 
has three regimes: (i) for low $l$ (i.e., $l < 2 R_S c$ for all $\theta$ at 
large radii), a torus does not form and $\MDOT_a= \MDOT_B$,
(ii) for intermediate $l$, or more appropriately where there is a narrow range 
of $\theta$ for which $l>2 R_S c$ so that $\theta_o > \theta_t$, 
$\MDOT_a \sim const$ and (iii) for  high $l$, or in the case for which 
$l> 2 R_S c$ at nearly all $\theta$ so that $\theta_o < \theta_t$,  
$\MDOT_a$  decreases with increasing $l_0$. Our limited data suggest that in 
the second regime, the actual value of the constant is dependent on the ratio 
$R_S/R_B$  for large $R_S/R_B$ but becomes independent on the ratio for small 
$R_S/R_B$ (i.e., for $R_S/R_B\simless 10^{-3}$). We conclude that 
the inclusion of even slow rotational motion of the inviscid flow 
at large radii can significantly
reduce $\MDOT_a$ compared to the Bondi rate, as the $\MDOT_a$ $vs.$ $l$ 
relation in the third regime indicates. For $R_S/R_B \ge 10^{-3.5}$, our 
results show that $\MDOT_a$ can be reduced by $\sim 1.5$ orders of magnitude 
compared to the Bondi rate. To reduce the mass accretion rate more, 
the accretion funnel needs to be much narrower than the  funnels allowed 
by our assumption of low angular momentum (eq. 27) for $R_S/R_B \ge 10^{-3.5}$.
It remains to be seen whether simulations for values of $R_S/R_B$ as low as 
those observed in astrophysical systems (i.e., $R_S/R_B \simless 10^{-6}$) 
will confirm our predictions  that the accretion funnel can be very narrow and 
subsequently the mass accretion rate very small. 

We note that the discontinuous change from quasi-spherical to 
a disk-like flow was first found by  Abramowicz  \& Zurek (1981).
Abramowicz \& Zurek studied analytically the adiabatic accretion
of a radial flow onto a black hole using the PW potential.
They considered  the case of constant specific angular momentum
and also found that for small $l$, the flow is quasi-spherical and
becomes transonic at $x_s \gg R'_S$ while for sufficiently
large $l$, the flow has a disk-like pattern and $x_s \leq 3 R'_S$.
Applying Abramowicz  \& Zurek's solution for the sonic point, we find
very good agreement between the location of the sonic
point on the equator and their prediction.
In particular, for $n\equiv1/(\gamma-1)=3/2$, $l=2 R_Sc$, 
and zero total energy, eq. 3.5 in Abramowicz \& Zurek (1981) yields that 
the sonic point equals $1+\sqrt{3}$, which is in a good  agreement with 
our result for $\theta=90^\circ$ where the sonic radius
is $\approx 2.5 R_S$ (e.g., see Figure~4). Our simulations
also agree with Abramowicz  \& Zurek's result that
the transition between 
quasi-spherical and disk-like accretion occurs at 
$l\approx 2 R_S c$.

We find that the shape of the polar funnel through which matter is accreted 
is partially constrained by the torus, which is made of matter that cannot 
accrete. The torus consists of material with roughly uniform angular momentum
and its structure does not depend much on the outer conditions. The accretion 
funnel is therefore also not very sensitive to the outer conditions in 
the limit that it is broad at large radii (i.e., large $\theta_o$). This is 
the key reason for the insensitivity of $\MDOT_a$ to $l$ in the second regime  
and its weak sensitivity in the third regime: the mass accretion rate depends 
on the geometry of the sonic surface at radii where the presence of the torus 
is important, and not only on the geometry at the outer boundary. Thus, 
the accretion flow in the funnel should depend on the physics that controls 
the flow in the torus. In particular, the introduction of energy dissipation,
and the transport of energy and angular momentum in the torus, may change
the shape of the torus and its effect on the polar regions, where material 
can accrete without the transport of angular momentum. One could argue that 
the total mass accretion rate (via both the funnel and torus) onto the black 
hole should increase when transport of  angular momentum and consequently 
accretion via the torus are allowed. However, it is not clear by how much 
$\MDOT_a$ will increase, if at all, because the energy and angular momentum 
from the torus are likely to be deposited in the polar region 
(see e.g., Stone, Pringle \& Begelman 1999; Blandford \& Begelman 1999; 
Blandford \& Begelman 2002a, 2002b; Hawley \& Balbus 2002). Indeed,  
outflow from the torus may interfere with the inflow in the funnel and 
$\MDOT_a$ may  well {\it decrease}. There is also a possibility that 
accretion via the funnel will decrease, not because of the dynamical effects 
due to outflowing material but because a torus in which energy is dissipated 
should be hotter and  thicker than a torus in which energy cannot be 
dissipated. On the other hand, it is not clear how accretion via the torus 
will change when supersonic accretion in the funnel is present. Our simulations
show that the range of $\theta$ that is occupied by the accretion funnel 
increases with decreasing distance from the black hole. In particular, 
accretion through the inner radius occurs for the entire range of $\theta$. 
We expect that the presence of a supersonic accretion flow near the equator 
at small radii may cause accretion via the torus to be less effective compared 
to the situation in which the material in the polar region is static. 
Most simulations of the formation of the torus have assumed that the material 
near the poles, outside the initially hydrostatic rotating torus, is static
or subsonic, and therefore unimportant dynamically. 

Even within our simple framework of inviscid flow, we anticipate that our 
results may change if we relax some of our assumptions about the geometry at 
the outer boundary. The assumptions we adopted are as simple as possible 
because we do not know the magnitude of the angular momentum of  material
in the environment of SMBHs, nor the angular distribution of $l$. 
In particular, it is rather unlikely that the $l$ distribution is axisymmetric.
Therefore, fully 3D calculations are required to explore how $\MDOT_a$ responds
to non-axisymmetric $l$ distributions. In the context of inviscid 
HD calculations similar to ours, one would expect that 3D effects 
may reduce $\MDOT_a$ compared to 2D axisymmetric calculations. For example, 
it is possible that  rotation at large radii occurs not just around one axis 
but around two or more axes. It is also possible that  rotation occurs around 
an axis that changes with $r$. Additionally it is plausible that there are 
``winds'' flowing past the SMBH or that the entropy in the SMBH environment 
is variable. In such a case,  material with too high $l$ to be accreted may
occupy nearly the entire range of $\theta$ at small radii (near or inside 
the sonic radius) and prevent low-$l$ material from accreting. In terms of 
accretion funnels, this would correspond to the situation in which the `funnel'
is very narrow and maybe not lined up with a radial vector.

Perhaps the best studied  massive black hole with a very low luminosity 
is the black hole in the Galactic center. Many models and ideas for
how to explain very low SMBH luminosities  have been explored
in the context of  Sgr~$\rm{A^\ast}$. In particular,
Melia (1992; 1994) proposed a spherical accretion model in which
the accretion flow is assumed to be in free-fall until a Keplerian
disk is formed within a small circularization radius. Coker \& Melia (2000)
looked at the problem of spherical accretion but with
the magnetic field being in subequipartition. The latter results
in a reduced bremsstrahlung emissivity and can help to explain
the low luminosity of Sgr~$\rm{A^\ast}$. A relatively small distance
to Sgr~$\rm{A^\ast}$ allows us to map the vicinity of the black hole
at the Galactic center. The complexity of such maps (e.g., in radio)
motivate three dimensional HD simulations.
For example, Coker \& Melia (1997) 
performed three-dimensional simulations of Bondi-Hoyle accretion
of  stellar winds onto a black hole. 
Clearly Sgr~$\rm{A^\ast}$
shows us that accretion onto black holes is a complex phenomenon
and the Bondi accretion formula should be used with great caution
because the assumption of spherical accretion is most certainly 
an oversimplification.

We finish with the observation that after three decades of studying
the Bondi problem with slight rotation at large radii 
(see, e.g., Henriksen \& Heaton 1975, 
Lynden-Bell 1978; Cassen \& Pettibone 1976; Sparke \& Shu 1980; Sparke 1982;
Abramowicz \& Zurek 1981, for analytic attempts to solve this
problem and see references in Section~1 for 2D and 3D numerical simulations),
we still find new complexities in the behavior of  accretion
flows. In our next phase of studying these flows, we will consider
3D MHD models  of radiatively inefficient flows.

ACKNOWLEDGMENTS: We thank J.M. Stone for useful discussions.
We acknowledge partial support from NSF
grant AST-9876887. DP also acknowledges partial support from NASA
grant NAG5-11736.
Some computations were  performed at Imperial College Parallel Computing
Center.

\newpage
\section*{ REFERENCES}
 \everypar=
   {\hangafter=1 \hangindent=.5in}

{
  Abramowicz, M.A., Chen, X., Kato, S., Lasota, J.-P., Regev, O. 
  1995, ApJ, 438, L37

  Abramowicz, M.A., Igumenshchev, I. V., Quataert, E., \& Narayan, R. 2002,
  ApJ, 565, 1101

  Abramowicz, M.A., \& Zurek, W.H. 1981, ApJ, 246, 314
  
  Baganoff, et at. 2001, ApJ, submitted (astro-ph/0102151)

  Balbus, S.A., \& Hawley, J.F. 1998, Rev. Mod. Phys., 70, 1.

  Balbus, S.A., \& Hawley, J.F. 2002, ApJ, 573, 749

  Begelman, M.C., 1978, A\&A, 70, 583

  Begelman, M.C., \& Meier, D.L. 1982, ApJ, 253, 873

  Blandford, R.D., \& Begelman, M.C. 1999, MNRAS, 303, L1
 
  Blandford, R.D., \& Begelman, M.C. 2002a, in preparation

  Blandford, R.D., \& Begelman, M.C. 2002b, in preparation
  
  Bondi, H. 1952, MNRAS, 112, 195
 
  Cassen, P., \& Pettibone, D. 1976, ApJ, 208, 500
 
  Chen, X., Taam, R.E., Abramowicz, M.A., \& Igumenshchev, I.V. 1997, MNRAS, 285, 439

  Clarke, D., Karpik, S., \& Henriksen, R.N. 1985, ApJS, 58, 81

  Coker, R.F., \& Melia, F. 1997, ApJ, 488, L149

  Coker, R.F., \& Melia, F. 2000, ApJ, 534, 723

  Di Matteo, T., Allen, S.W.,  Fabian, A.C., Wilson, A.S., \& Young, A.J. 2002,
  ApJ, submitted (astro-ph/0202238)

  Di Matteo, T., Carilli, C.L., \& Fabian, A.C. 2001, ApJ, 547, 731

  Di Matteo, T., Fabian, A.C., Rees, M.J., Carilli, C. L., \& Ivison, R.J. 
  1999, MNRAS, 305, 492

  Di Matteo, T., Quataert, E., Allen, S.W., Narayan, R., \& Fabian, A. C.
  2000 MNRAS, 311, 507 

  Hawley, J.F. 1986, in Radiation Hydrodynamics in Stars and Compact Objects, ed.  D. Mihalas \& K.-H. Winkler (New York:Springer), 369
 
  Hawley, J.F., \& Balbus, S.A.  2002, ApJ, 573, 749

  Hawley, J.F., Balbus, S.A., \& Stone, J.M. 2001, ApJ, 554, L49

  Hawley, J.F., Smarr, L.L., \& Wilson, J.R. 1984a, ApJ, 277, 296
 
  Hawley, J.F., Smarr, L.L., \& Wilson, J.R. 1984b, ApJS, 55, 211

  Henriksen, R.N., \& Heaton, K.C. 1975, MNRAS, 171, 27
 
  Ichimaru, S. 1977, ApJ, 214, 840.

  Igumenshchev, I.V., \& Abramowicz, M.A. 1999, MNRAS, 303, 309
 
  Igumenshchev, I.V. \& Narayan, R. 2002, ApJ, 566, 137

  Kryukov, I.A., Pogorelov, N.V., Bisnovatyi-Kogan, G.S., Anzer, U., \&
  B\"{o}rner, G. 2000, A\&A, 364, 901

  Lynden-Bell, D. 1978, PhyS, 17, 185

  Loewenstein, M., Mushotzky, R.F., Angelini, L., Arnaud, K.A., \& 
  Quataert, E. 2001, ApJ, 555, L21

  Machida, M., Matsumoto, R., \& Mineshige, S. 2001, PASJ, 53, L1

  Melia, F. 1992, ApJ, 387, L25

  Melia, F. 1994, ApJ, 426, 577
   
  M\'{e}sz\'{a}ros, P. 1975, A\&A, 44, 59  
 
  Molteni, D., Lanzafame, G., \& Chakrabarti, S. 1994, ApJ, 425, 161

  Narayan, R., Igumenshchev, I.V., \& Abramowicz, M.A. 2000, ApJ, 539, 798

  Narayan, R., Quataert E., Igumenshchev, I.V., \& Abramowicz, M.A. 2002,
  ApJ, submitted  (astro-ph/0203026)
 
  Narayan, R., \& Yi, I. 1994, ApJ, 428, L13
 
  Narayan, R., \& Yi, I. 1995, ApJ, 444, 231
  
  Ostriker, J.P., McCray, R., Weaver, R., \& Yahil, A. 1976, ApJ, 208, L61
 
  Paczy\'{n}ski, B., \& Wiita, J. 1980, A\&A, 88, 23

  Quataert, E., \& Gruzinov A. 2000, ApJ, 545, 842
 
  Quataert, E., \& Narayan R. 1999, ApJ, 520, 298

  Rees, M.J., Begelman, M.C., Blandford, R.D., \& Phinney, E.S. 1982, 
  Nature, 295, 17

  Ruffert, M. 1994, ApJ, 427, 342
 
  Ryu, D., Brown, G.L, Ostriker, J.P., \& Loeb, A. 1995, ApJ, 452, 364

  Shakura, N.I., \& Sunyaev, R.A. 1973, A\&A, 24, 337

  Shapiro, S.L. 1973, ApJ, 180, 531  

  Sparke, L.S. 1982, ApJ, 254, 456

  Sparke, L.S., \& Shu, F.H. 1980, ApJ, 241, L65
 
  Stone, J.M., \& Norman, M.L. 1992, ApJS, 80, 753

  Stone, J.M., \& Pringle, J.E. 2001, MNRAS, 322, 461

  Stone, J.M., Pringle, J.E., \& Begelman, M.C. 1999, MNRAS, 310, 1002

  Toropin, Y.M., Toropina, O.D., Savelyev, V.V., Romanova, M.M., 
  Chechetkin, V.M., \& Lovelace, R.V.E. 1999, ApJ, 517, 906

}

\newpage

\begin{table*}
\footnotesize
\begin{center}
\caption{ Summary of parameter survey.}
\begin{tabular}{l c  c c c c c c c c l  } \\ \hline 
         &       &    &            &             &          &   & & &  \\
Run$^\ast$ & Resolution   & $R'_S$ & $R'_C$  & $l_0$ & $\theta_o$ & $f(\theta)$  & $t_f$     & $l^{max}_a$  & $\MDOT_a/\MDOT_B$ & Comments  \\ 
         &                &   &            &     &       & & & & &  \\

A02f1a  &   100        & $10^{-2}$    & $2\times10^{-1}$& 1.6 & $68^\circ$ & $1-|\cos{\theta}|$   & 35 & $0.80$ &   0.57 &    \\   
A03f1a  &   100        & $10^{-2}$    & $5\times10^{-1}$& 2.5 & $53^\circ$ & $1-|\cos{\theta}|$   & 36 & $0.80$ &   0.40 &    \\   
A04f1a  &   100        & $10^{-2}$    & $1\times10^{0}$ & 3.5 & $44^\circ$ & $1-|\cos{\theta}|$   & 12 & $0.81$ &   0.50 &    \\   
A08f1a  &   100        & $10^{-2}$    & $1\times10^{2}$ & 35.3& $14^\circ$ & $1-|\cos{\theta}|$   & 36 & $0.87$ &   0.08 &    \\   

         &                &   &            &            & & & & & &  \\
A03f1b  &   140        & $10^{-2}$    & $5\times10^{-1}$& 2.5 & $53^\circ$ & $1-|\cos{\theta}|$   & 441&  $0.40$ &   0.97 & $r_o =12 R_B$   \\   
         &                &   &            &            & & & & & &  \\
A07f2a  &   100        & $10^{-2}$    & $5\times10^{-1}$& 2.5 & $18^\circ$ & $1-\cos^{10}{\theta}  $ & 31&  $0.82$ &   0.18 &    \\   

         &                &   &                     &   & & & & &  \\
B01f1a  &   140        & $10^{-3}$    & $8\times10^{-3}$& 1  & $90^\circ$ & $1-|\cos{\theta}|$   & 3.5 & $0.87$ &   0.30  &     \\   
B03f1a  &   140        & $10^{-3}$    & $5\times10^{-2}$& 2.5& $53^\circ$ & $1-|\cos{\theta}|$   & 5.9 & $0.87$ &   0.30 &    \\   
B04f1a  &   140        & $10^{-3}$    & $1\times10^{-1}$& 3.5& $44^\circ$ & $1-|\cos{\theta}|$   & 9.3&  $0.86$ &   0.30 &    \\   
B05f1a  &   140        & $10^{-3}$    & $5\times10^{-1}$& 7.9& $29^\circ$ & $1-|\cos{\theta}|$   & 4.4 & $0.87$ &   0.24 &    \\   
B06f1a  &   140        & $10^{-3}$    & $1\times10^{0}$& 11.2& $24^\circ$ & $1-|\cos{\theta}|$    & 4.3&  $0.87$ &   0.22 &    \\   
         &                &   &                     &   & & &  & &   \\
B04f1b  &   180        & $10^{-3}$    & $1\times10^{-1}$& 3.5& $44^\circ$ & $1-|\cos{\theta}|$   & 8.2&  $0.89$ &   0.30 &  $r_o=12R_B$  \\   

         &                &   &                     &   & & & & &   \\
B08f3a &   140        & $10^{-3}$    & $8\times10^{-3}$ & 1&$8.1^\circ$ & step function   &6.7 &  $0.88$ &   0.05 &  \\   
B09f3a &   140        & $10^{-3}$    & $8\times10^{-3}$ & 1&$6.3^\circ$ & step function   &8.0 &  $0.88$ &   0.04 &   \\   
B10f3a &   140        & $10^{-3}$    & $8\times10^{-3}$ & 1&$4.5^\circ$ & step function   &6.7 &  $0.88$ &   0.03 &   \\   

         &                &   &                     &   & & & & &   \\
B08f2a  &   140        & $10^{-3}$    & $1\times10^{-1}$& 3.5& $8.1^\circ$ & $1-\cos^{10}{\theta}  $ &6.7 &  $0.88$ &   0.11 &   \\   

         &                &   &                     &   & & & & &   \\
C04f1a  &   160        & $10^{-3.5}$    & $3.1\times10^{-2}$& 3.5& $44^\circ$ &
$1-|\cos{\theta}|$   & 3.6 & $0.8$ &     $0.20$  & \\   
         &                &   &                     &   & & & & &   \\
C08f2a  &   160        & $10^{-3.5}$    & $3.1\times10^{-2}$& 3.5& $8.1^\circ$ &
$1-\cos^{10}{\theta} $ & 2.7 &  $0.8$ &   0.09 &   \\   
\hline
\end{tabular}

$\ast$ We use the following convention to label our runs: the first character
in the name refers to $R'_S$, (i.e., A, B, and C are for $R'_S=10^{-2},
10^{-3}$, and $10^{-3.5}$, respectively). The second and third
characters refer
to $\theta_o$ (i.e., 01, 02, 03, 04, 05, 06, 07, 08, 09 and, 10 are for 
$\theta_o= 90^\circ, 68^\circ, 53^\circ, 44^\circ, 29^\circ, 24^\circ,
18^\circ, 14^\circ, 8.1^\circ, 6.3^\circ$, and $4.5^\circ$, respectively), 
the fourth and fifth characters refer to the angular
distribution of $l$ (i.e., $f1$, $f2$, and $f3$ are 
for $1-|\cos{\theta}|$, $1-\cos^{10}{\theta}$, 
and the step function (see eq. 25), respectively), 
and finally the sixth character refers to the
outer radius of the computational domain (i.e., a and b stands for
$r_o=1.2~R_B$ and $12~R_B$, respectively).

\end{center}
\normalsize
\end{table*}

\eject 

Fig. 1. A sequence of specific angular momentum contours (top)
and velocity fields (bottom) from run B04f1a at times 0.03, 0.06,
4.56, and 9.27.
The specific angular momentum is in units of $2 R_S c$. The minimum
of $l$ (contour closest to the rotational $z-$axis) is 0.3 and 
the contours levels are equally spaced at intervals of $\Delta l=0.3$.
Note that the maximum of $l$ is 3.6 at the beginning
of the simulation and 3.3 at the end of the simulation, indicating that
lower angular momentum material is displacing higher angular momentum
material in the torus.
Only the $l=0.9$
contour is labeled to show the boundary between the flow that does
and does not accrete.
We suppressed
the velocity in the innermost part of the flow to better show the flow
pattern at large radii.

Fig. 2. The specific angular momentum contours and velocity field
at the end of run B04f1a. This figure is an enlargement of the rightmost
panels from Figure~1. Note that at small radii the nonaccreting flow,
the torus, is of nearly constant specific angular momentum.
The ``figure-8'' contour on both panels marks the radial sonic surface (the
location where $M_r=v_r/c_s=1$).

Fig. 3. Quantities at the inner boundary in model B04f1a  at $t_f=9.27$.
The solid and dashed lines on the top panel show the radial
and latitudinal Mach numbers, respectively. The specific angular momentum
(middle panel) is in units of $2 R_S c$ while the mass flux
density (bottom panel) is in units of the Bondi mass flux density
[i.e., $-(\rho v_r)_B\equiv \MDOT_B/ (4 \pi r^2_i)$].

Fig. 4. Quantities on the sonic surface  in model B04f1a at $t_f= 9.27$.
The top panel shows the sonic radius as a function of the latitude
in units of $R_B$.
The second panel from the top  shows $l$ in units of $2 R_S c$ (solid line)
and  the latitudinal Mach number
(dashed line). The second panel from the bottom  shows the mass flux
density in units of the Bondi mass flux density at the radial sonic point.
[i.e., $-(\rho v_r)_B\equiv \MDOT_B / (4 \pi R^2_B x^2_s) $].
The bottom panel shows a measure of the deviation from radial 
flow, $\Delta \theta$, as defined in the text.

Fig. 5. The time evolution of the mass accretion rate in units of the
Bondi rate, for model B04f1a.

Fig. 6. The specific angular momentum contours
at the end of run B01f1a. The right-hand panel is  an enlargement of 
the left-hand  panel.
The ``figure-8'' contour on the right-hand  panel marks the 
the radial sonic surface (the location where $M_r=1$).
The spacing of the $l$ contours is as in Figure~1. Note that  
runs B04f1a (Fig. 1) and B01f1a have different $l$ distributions at large radii yet
at small radii the flows in both runs are qualitatively similar (compare
figure~2 and the right-hand panel here).

Fig. 7. The mass accretion rate as a function of $\theta_o$.
The angle $\theta_o$ is the polar angle at which $l=2 R_S c$ at the outer
boundary. The solid line represents results for all of our model
with $R_S/R_B=10^{-3}$, regardless of the angular distribution of $l$
and regardless of $l_0$ (see Table~1). The dashed line represent
the theoretical prediction that $\MDOT_a$ scale with the solid angle
within which $l< 2 R_S c$ at $R_B$.
The mass accretion
rate is in units of the corresponding Bondi rate.

\newpage

\begin{figure}
\begin{picture}(180,480)
\put(110,180){\includegraphics{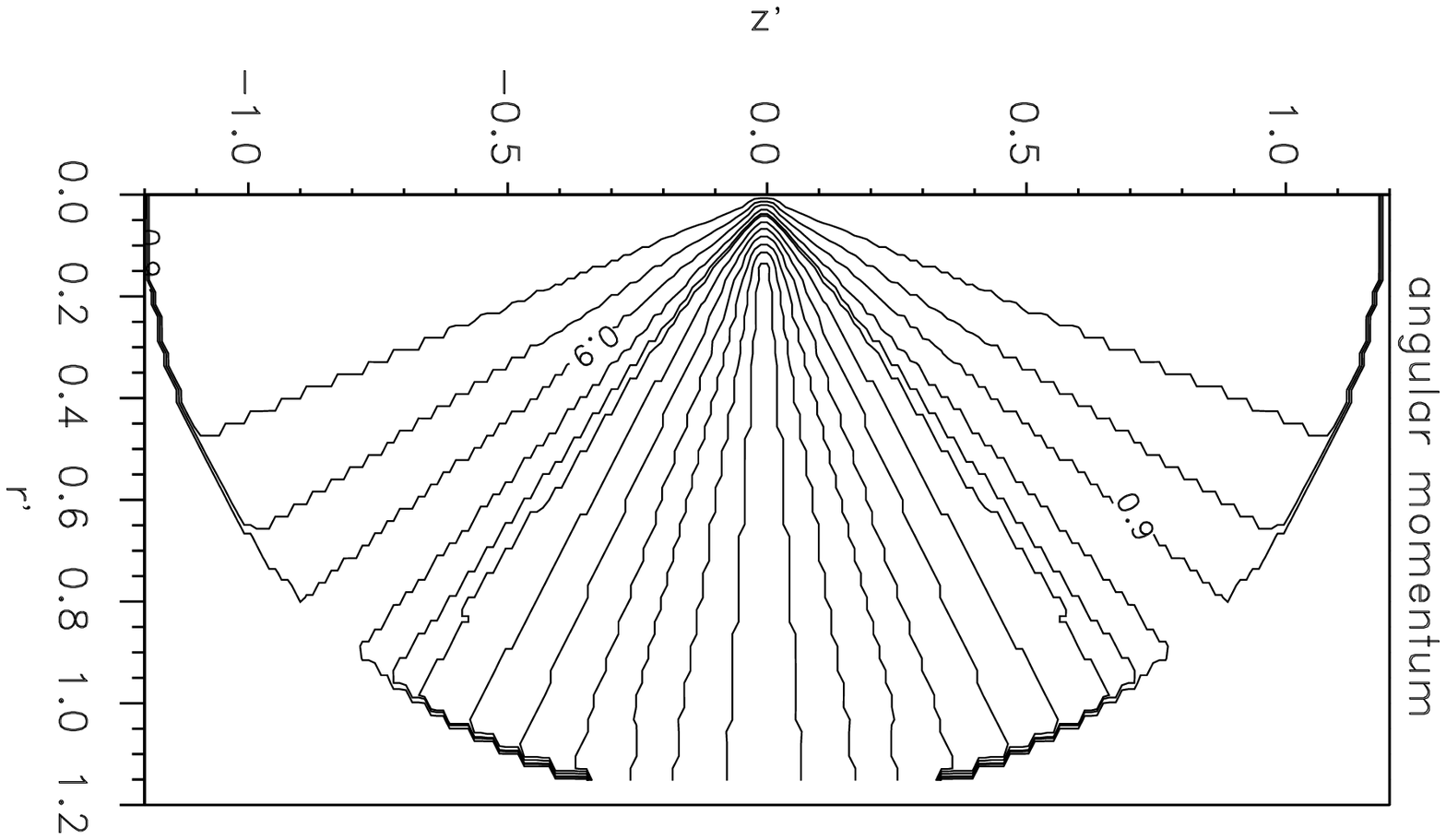}}
\put(110,50){\includegraphics{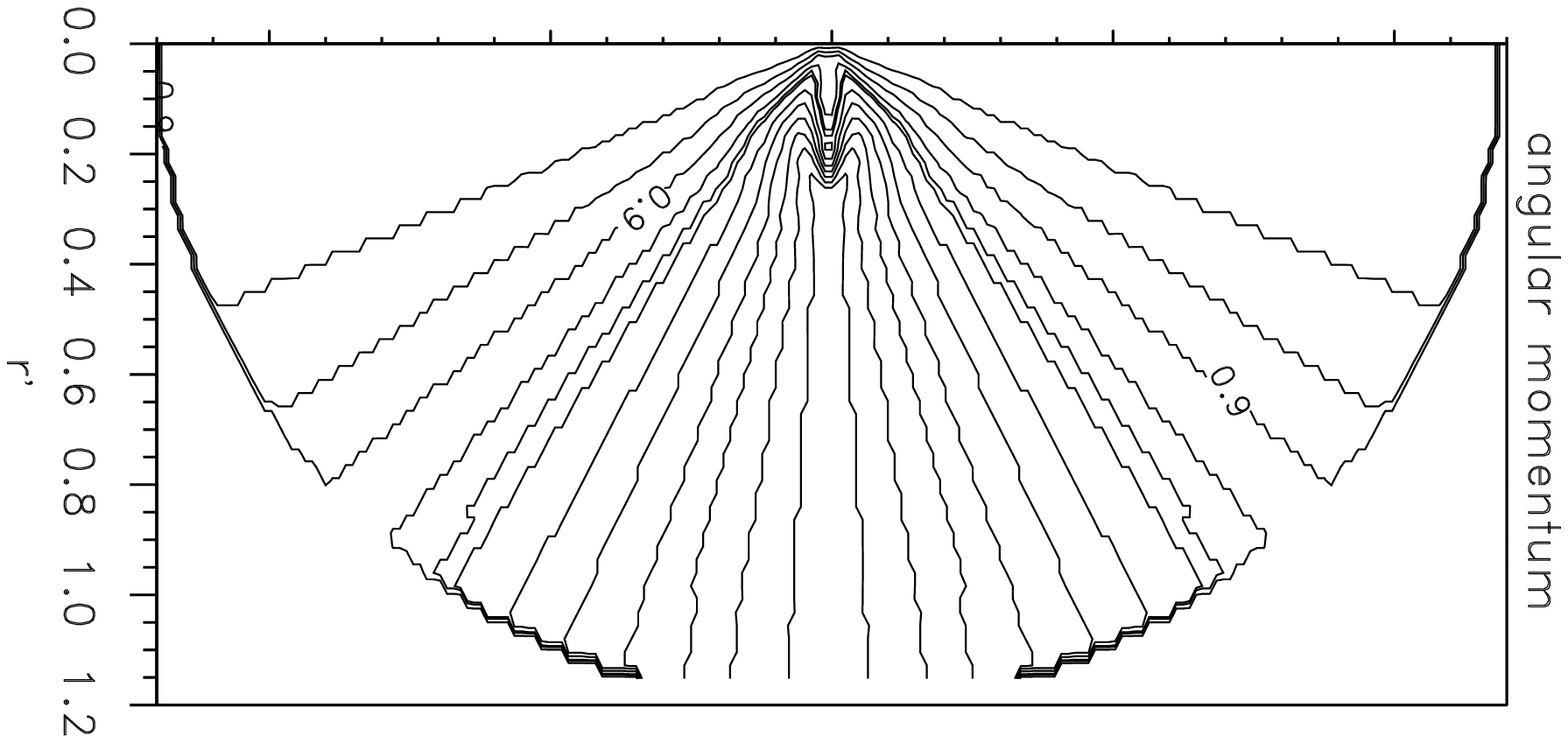}}
\put(110,-80){\includegraphics{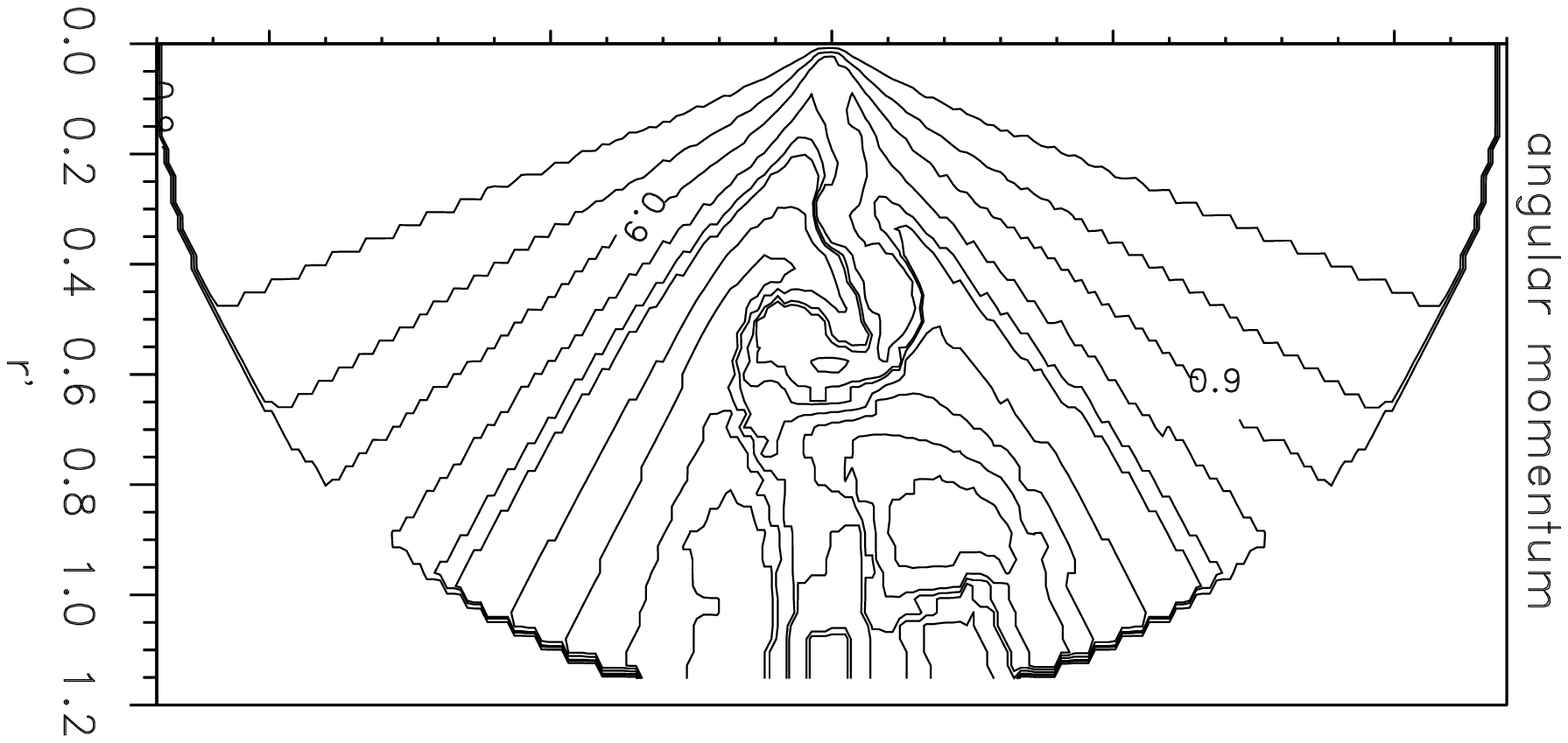}}
\put(110,-210){\includegraphics{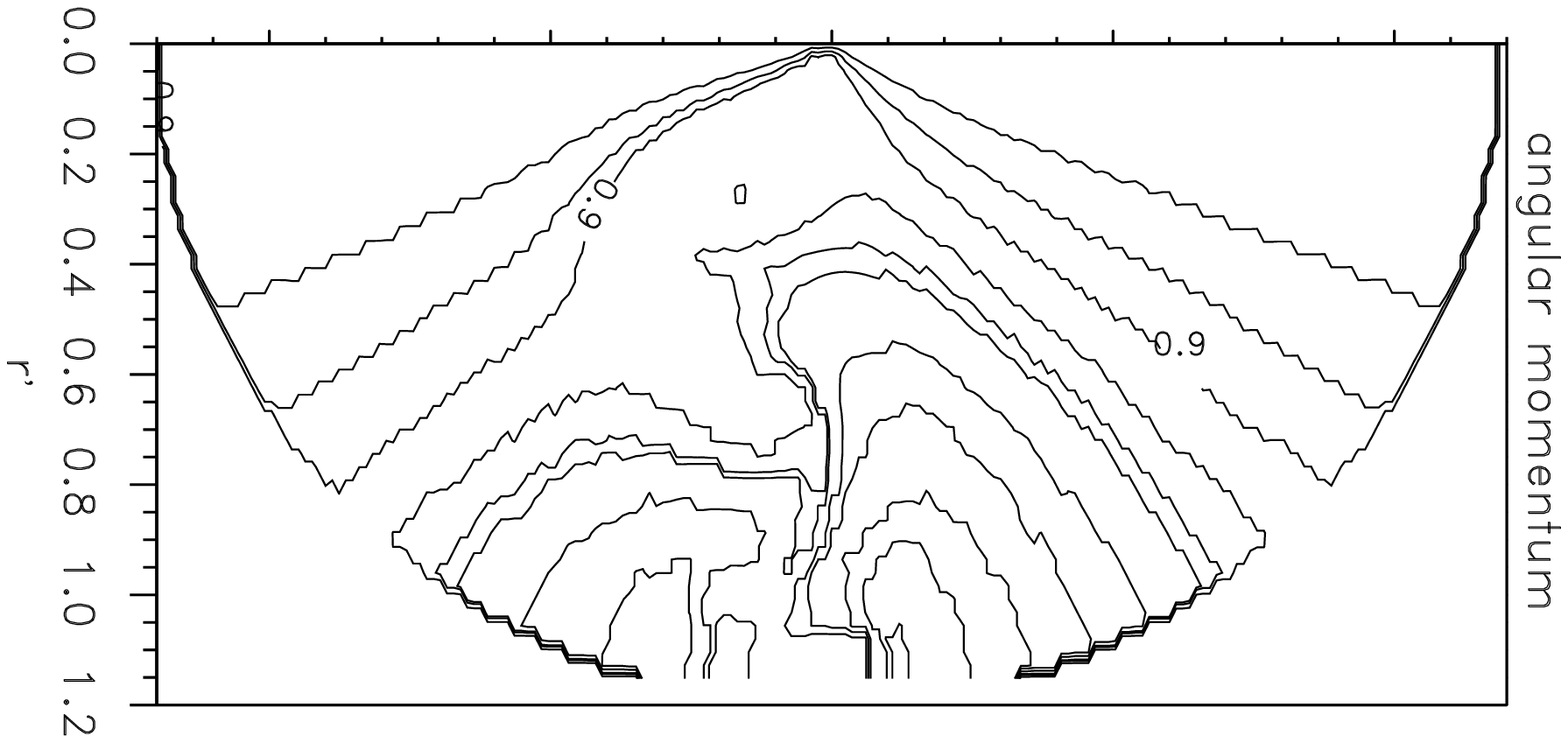}}

\put(-330,180){\includegraphics{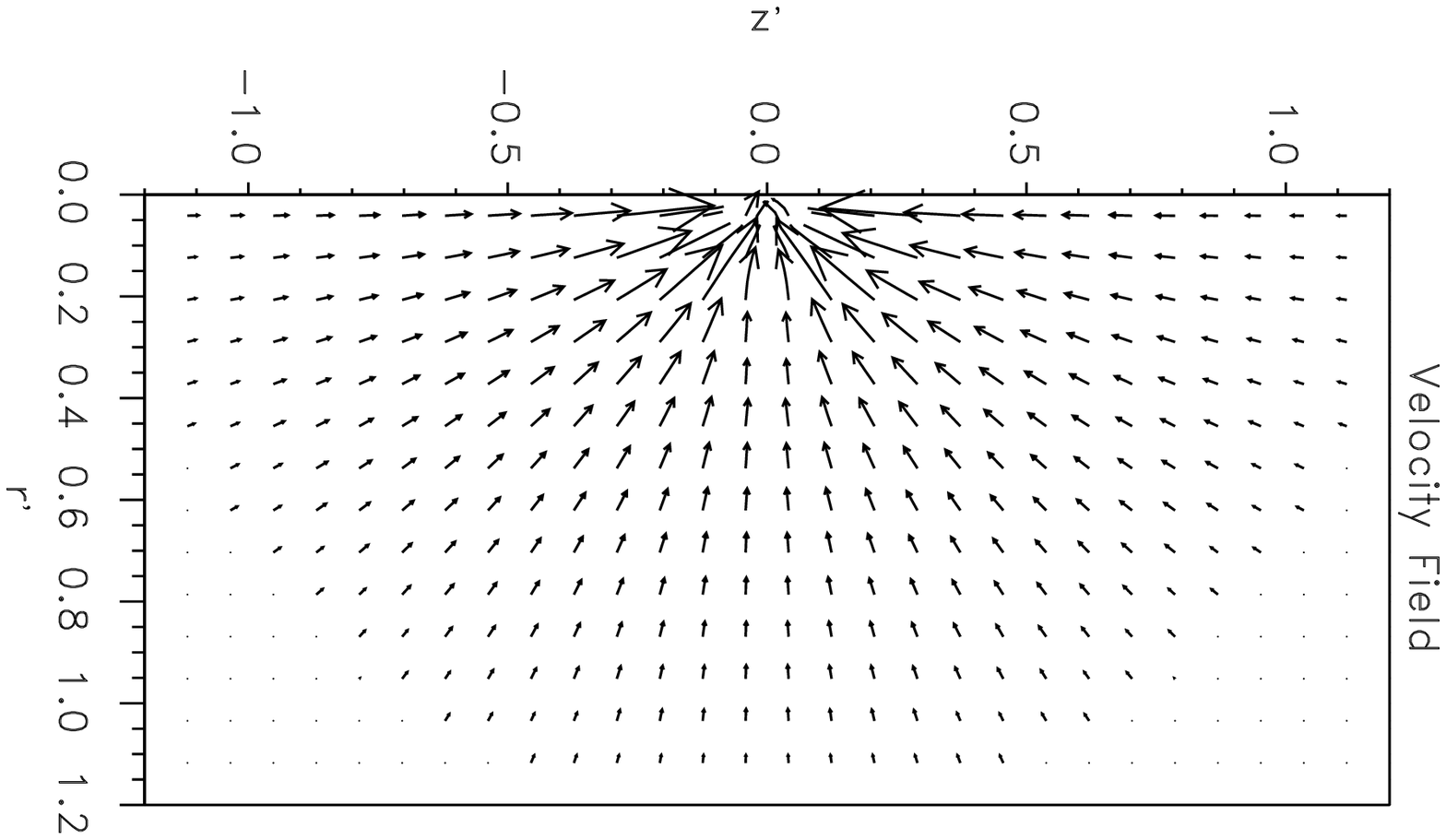}}
\put(-330,50){\includegraphics{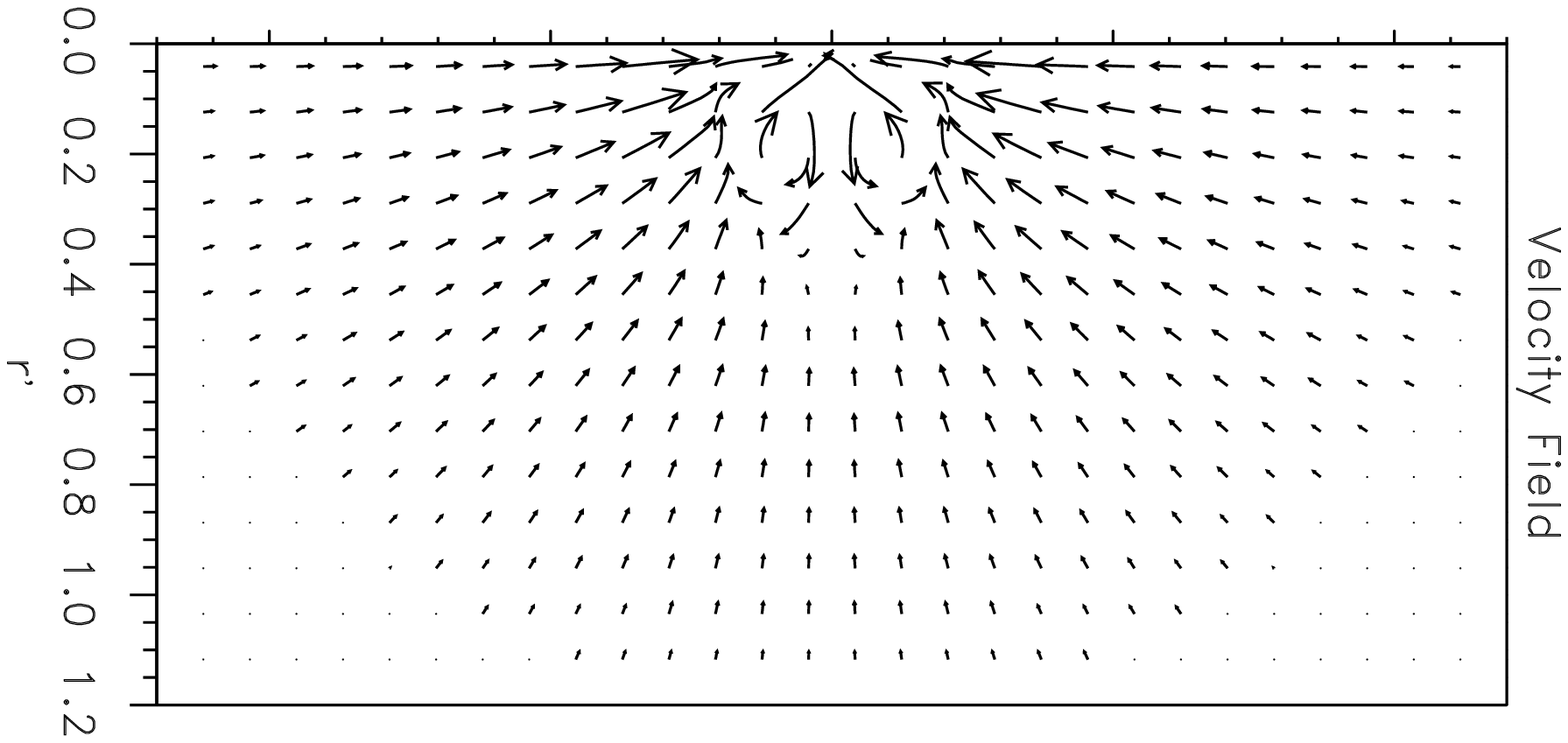}}
\put(-330,-80){\includegraphics{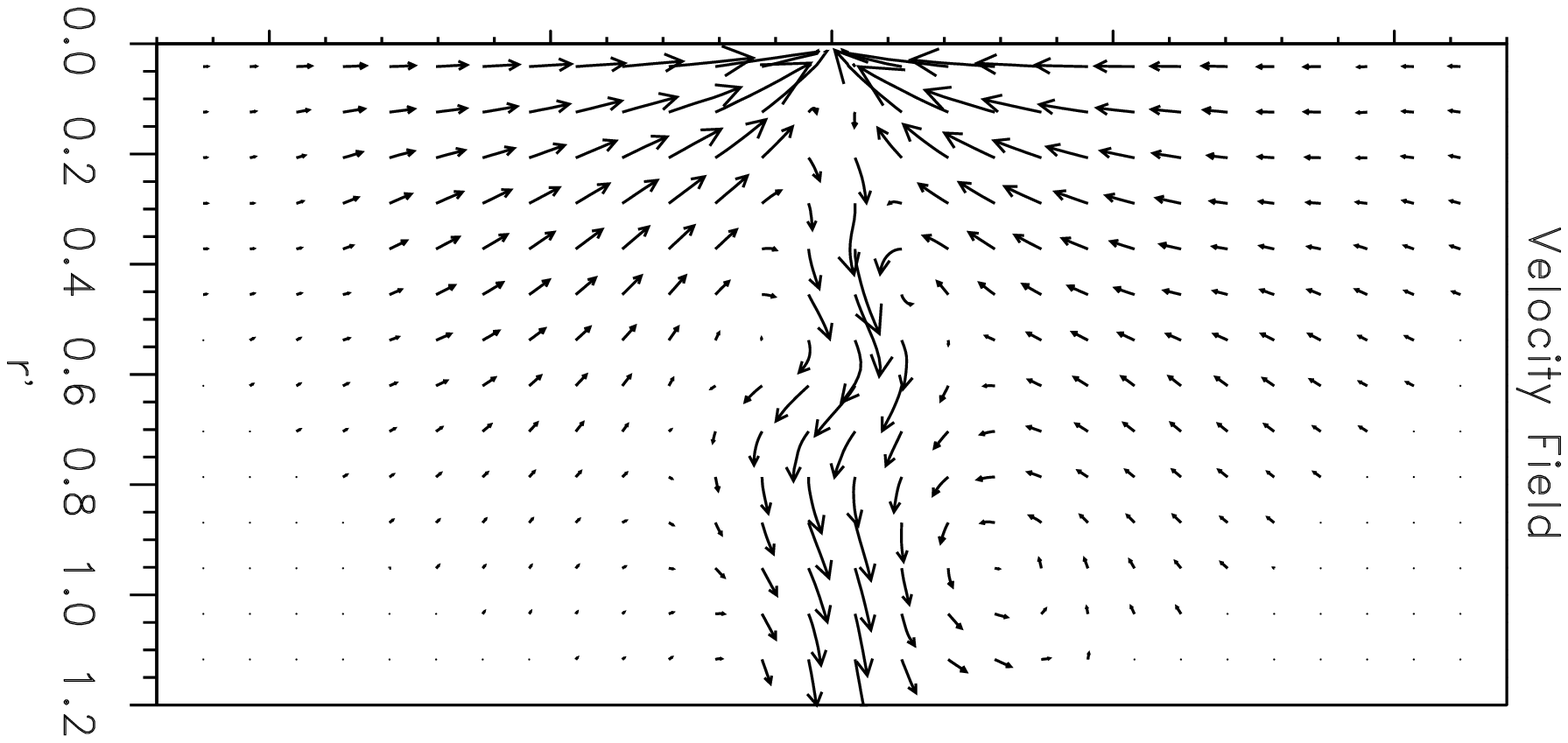}}
\put(-330,-210){\includegraphics{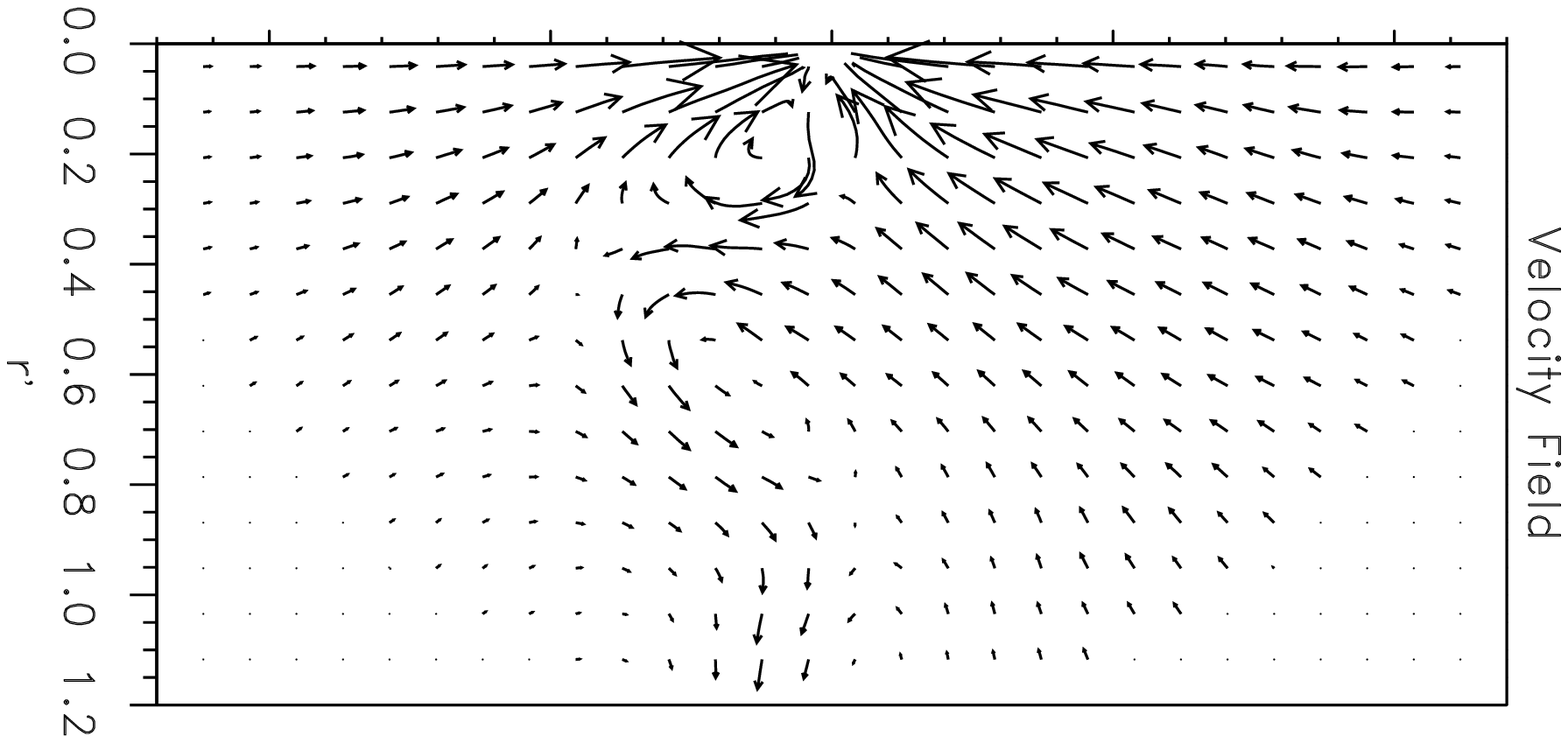}}
\end{picture}
\caption{ 
}
\end{figure}

\newpage

\begin{figure}
\begin{picture}(180,430)
\put(-90,0){\includegraphics{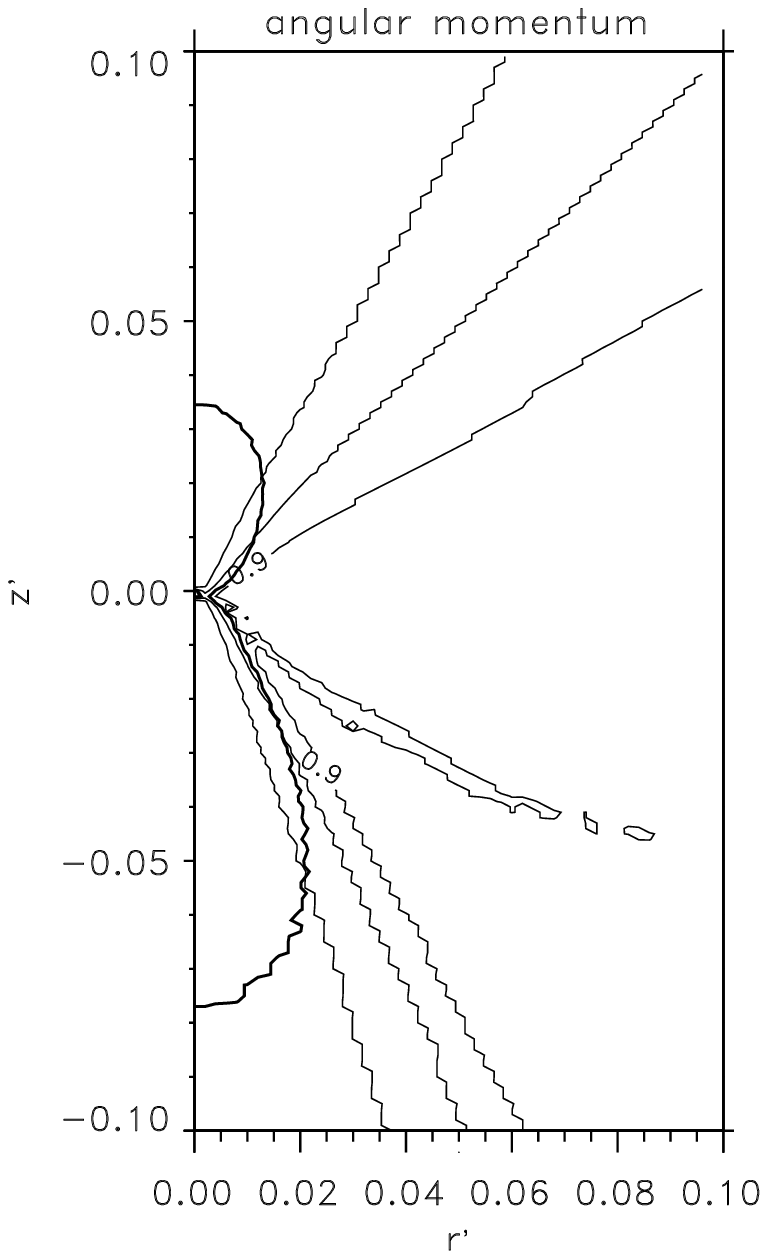}}

\put(40,0){\includegraphics{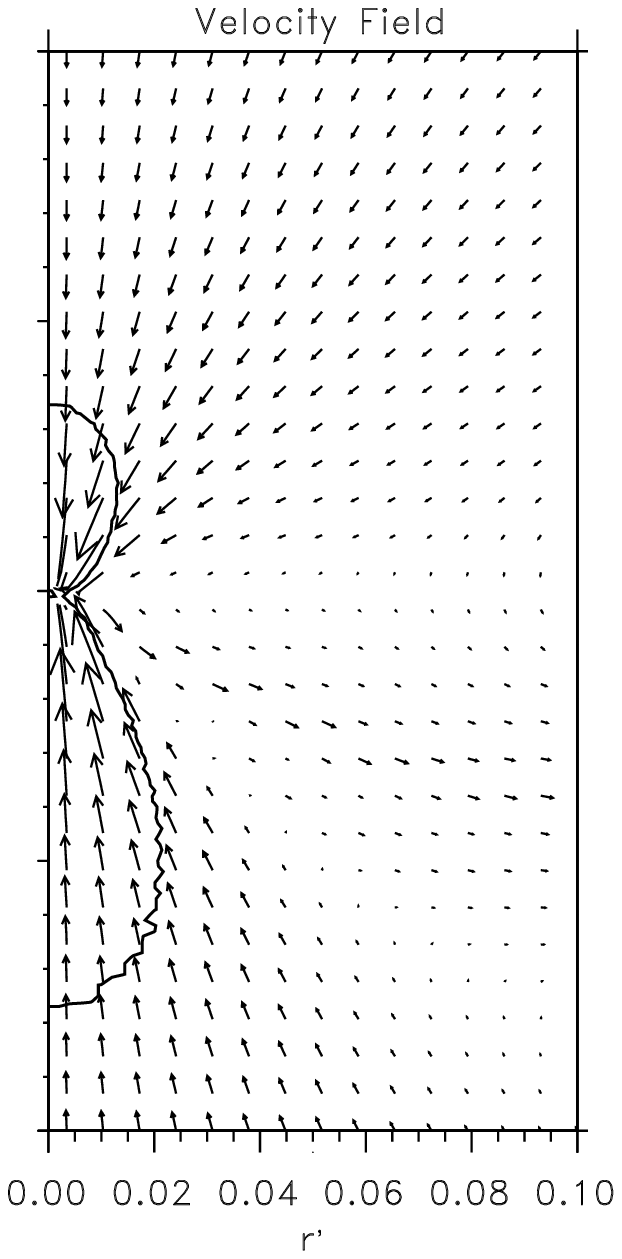}}
\end{picture}
\caption{ 
}
\end{figure}

\eject
\newpage

\begin{figure}
\begin{picture}(180,330)
\put(210,260){\includegraphics{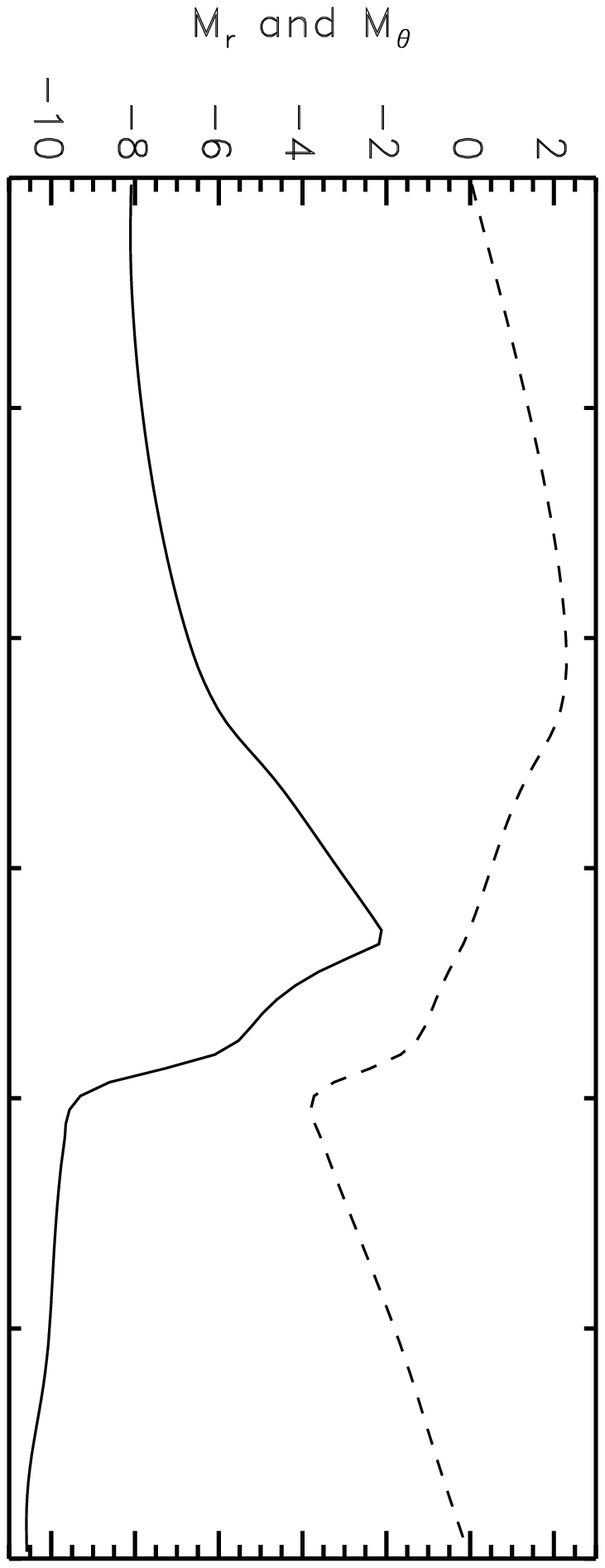}}
\put(210,180){\includegraphics{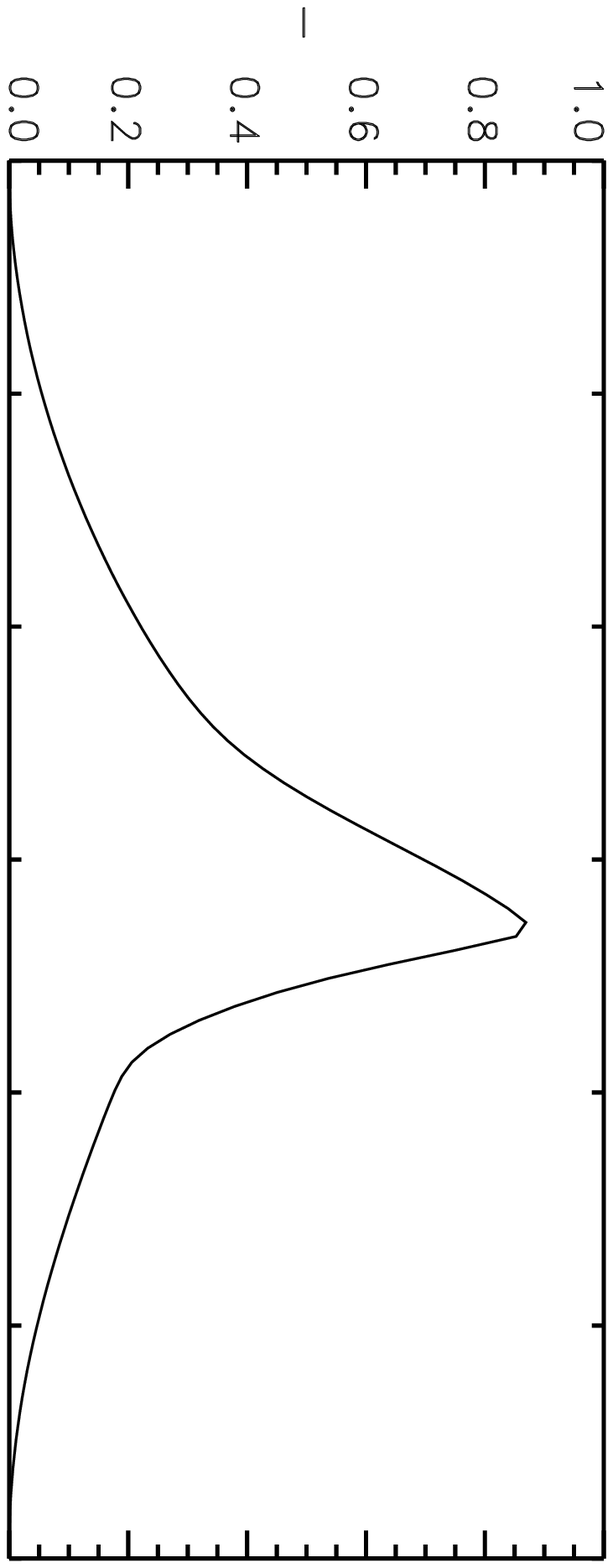}}
\put(210,100){\includegraphics{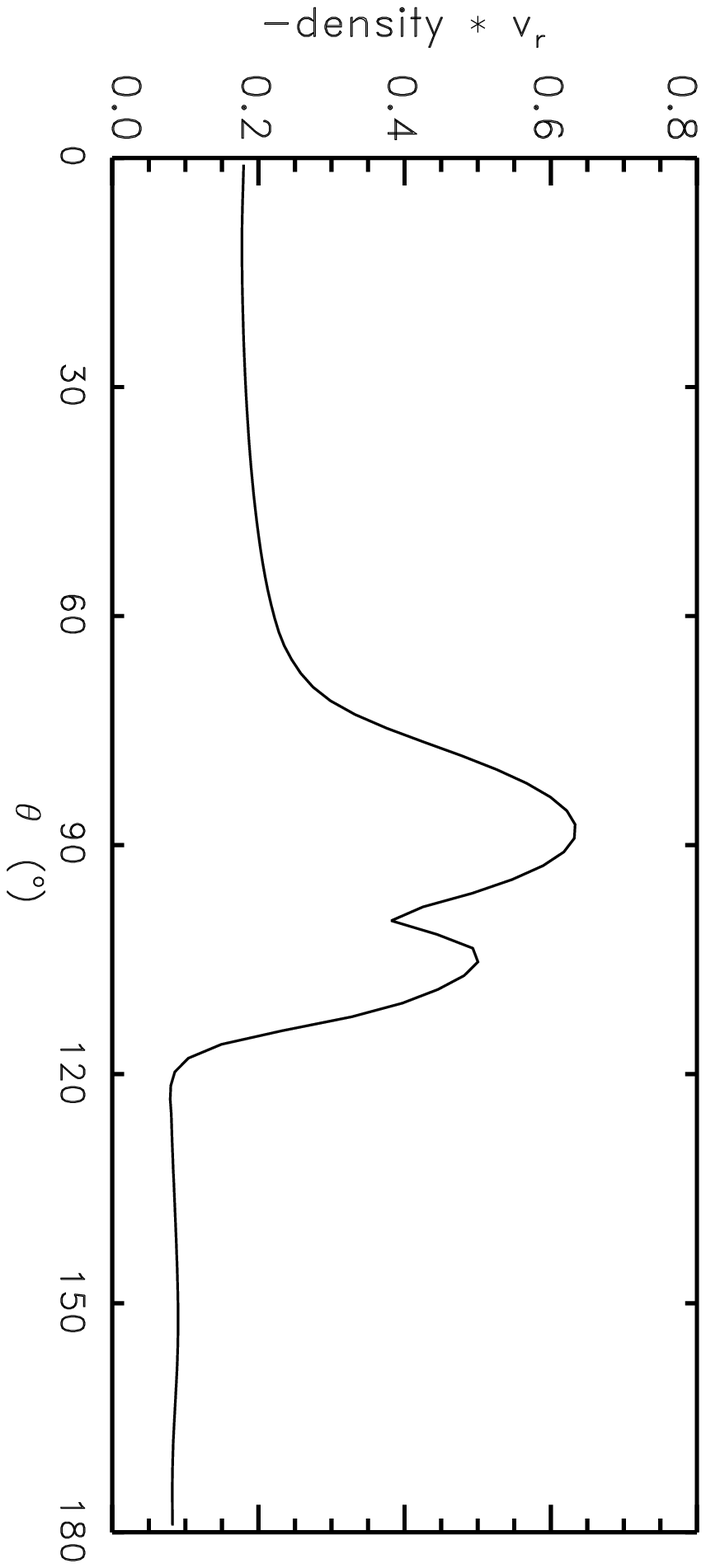}}
\end{picture}
\caption{
}
\end{figure}

\eject
\newpage

\begin{figure}
\begin{picture}(180,330)
\put(210,240){\includegraphics{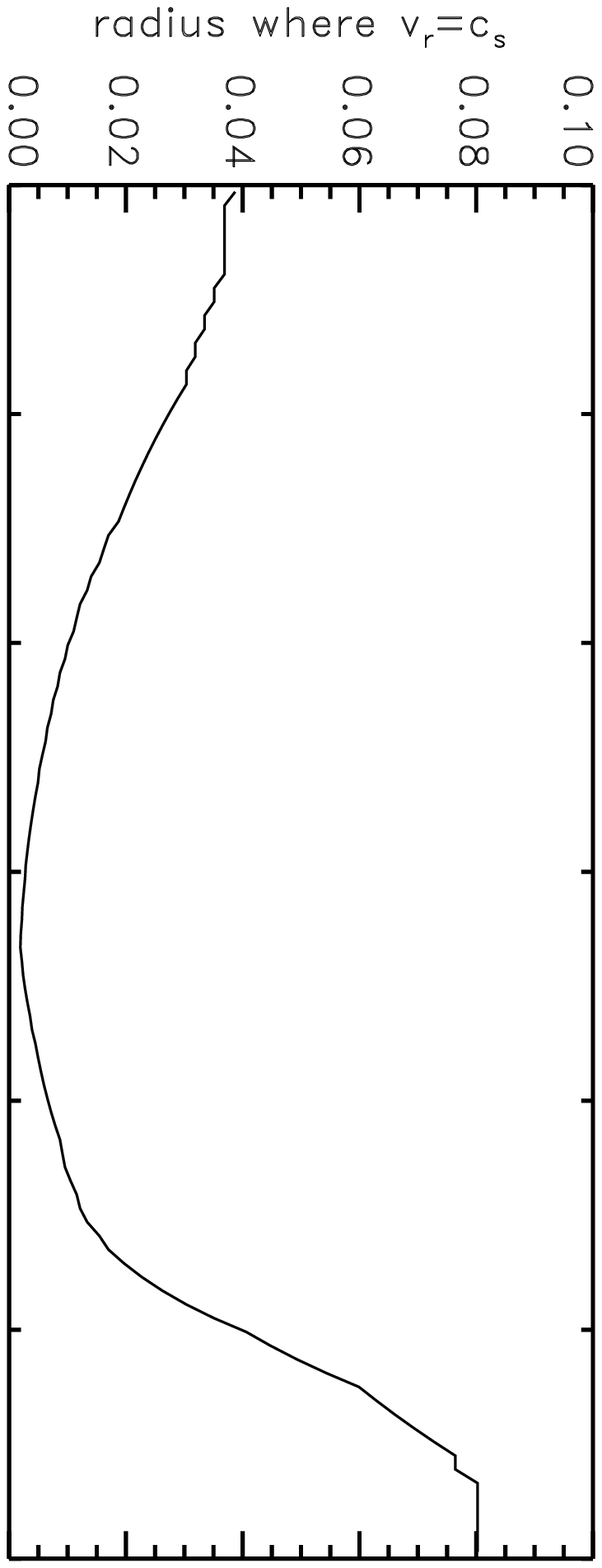}}
\put(210,160){\includegraphics{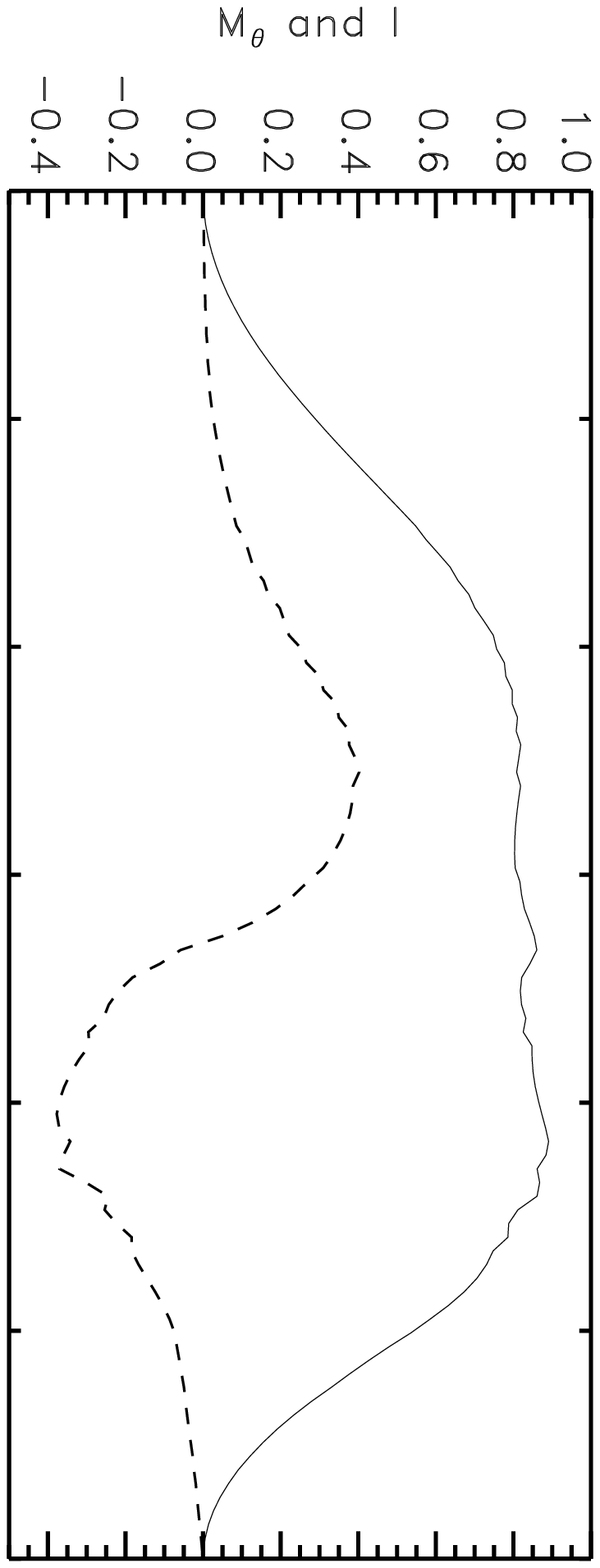}}
\put(210,80){\includegraphics{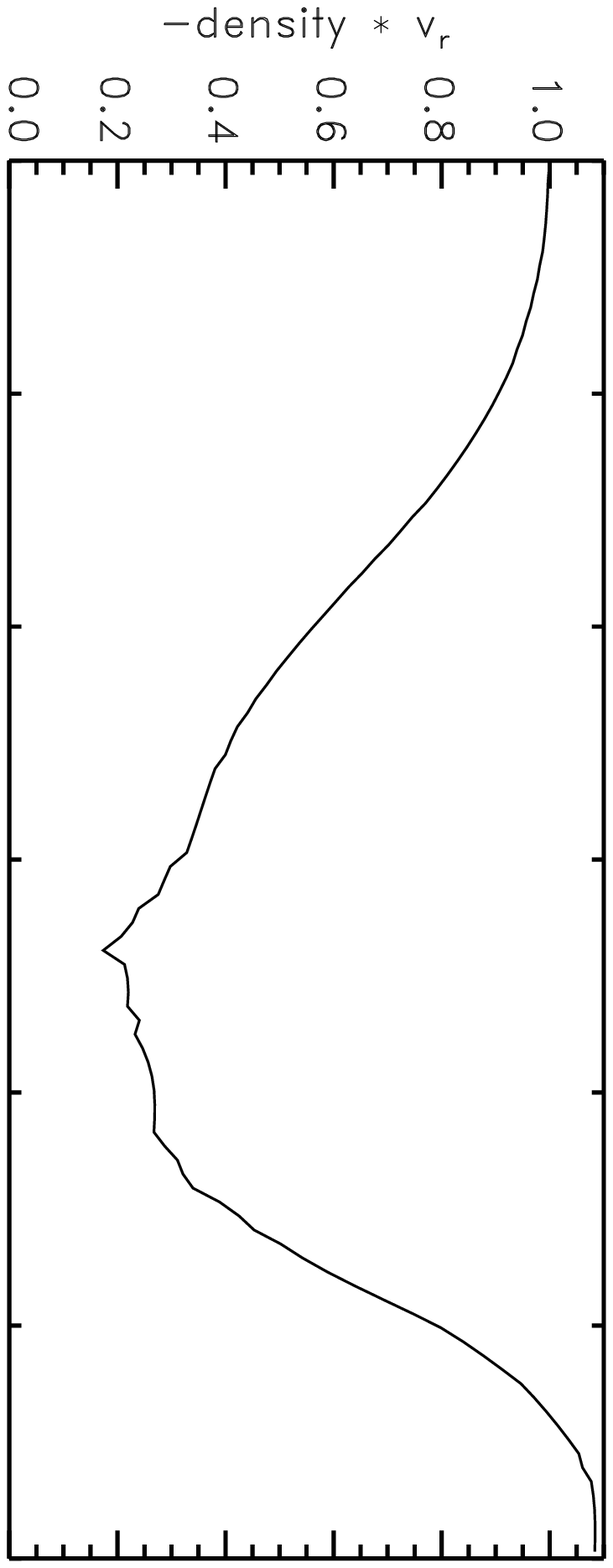}}
\put(210,0){\includegraphics{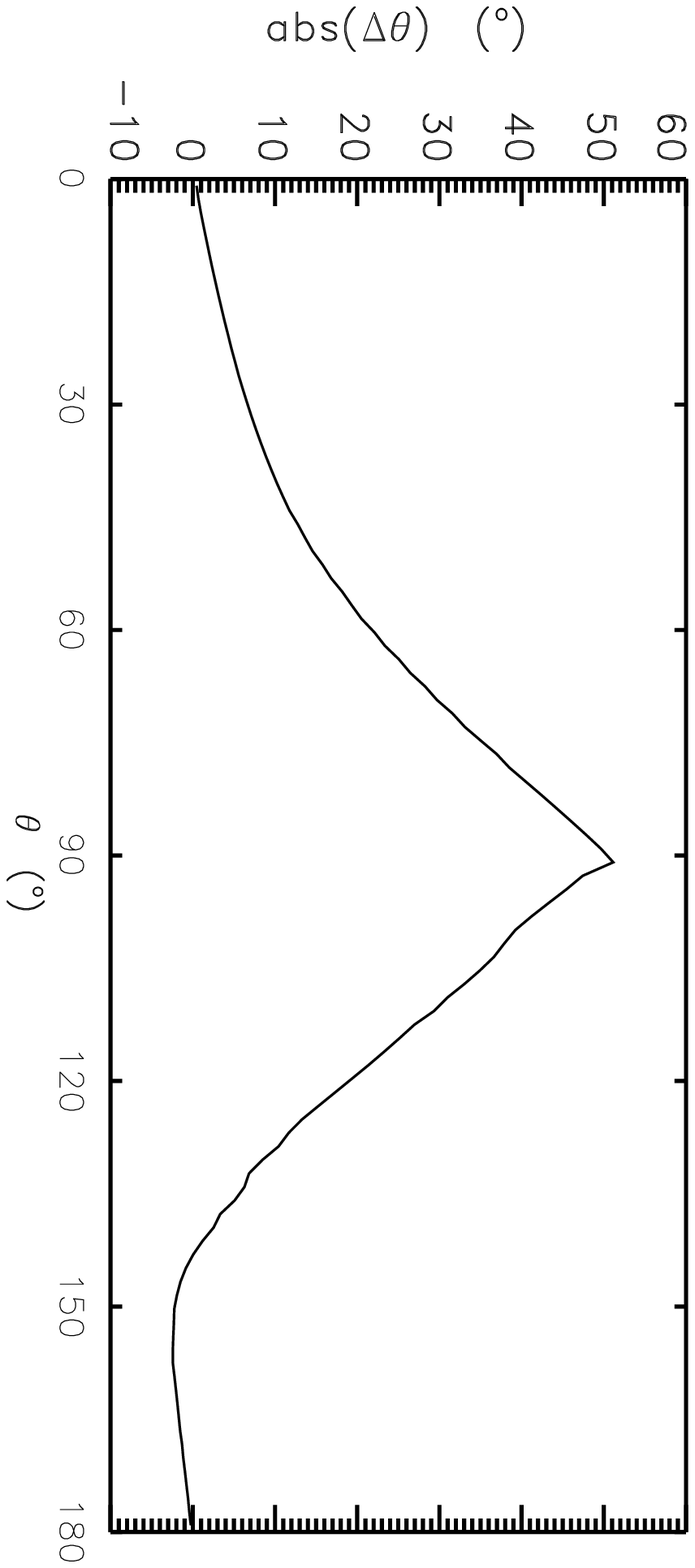}}
\end{picture}
\caption{
}
\end{figure}

\begin{figure}
\begin{picture}(180,400)
\put(0,0){\includegraphics{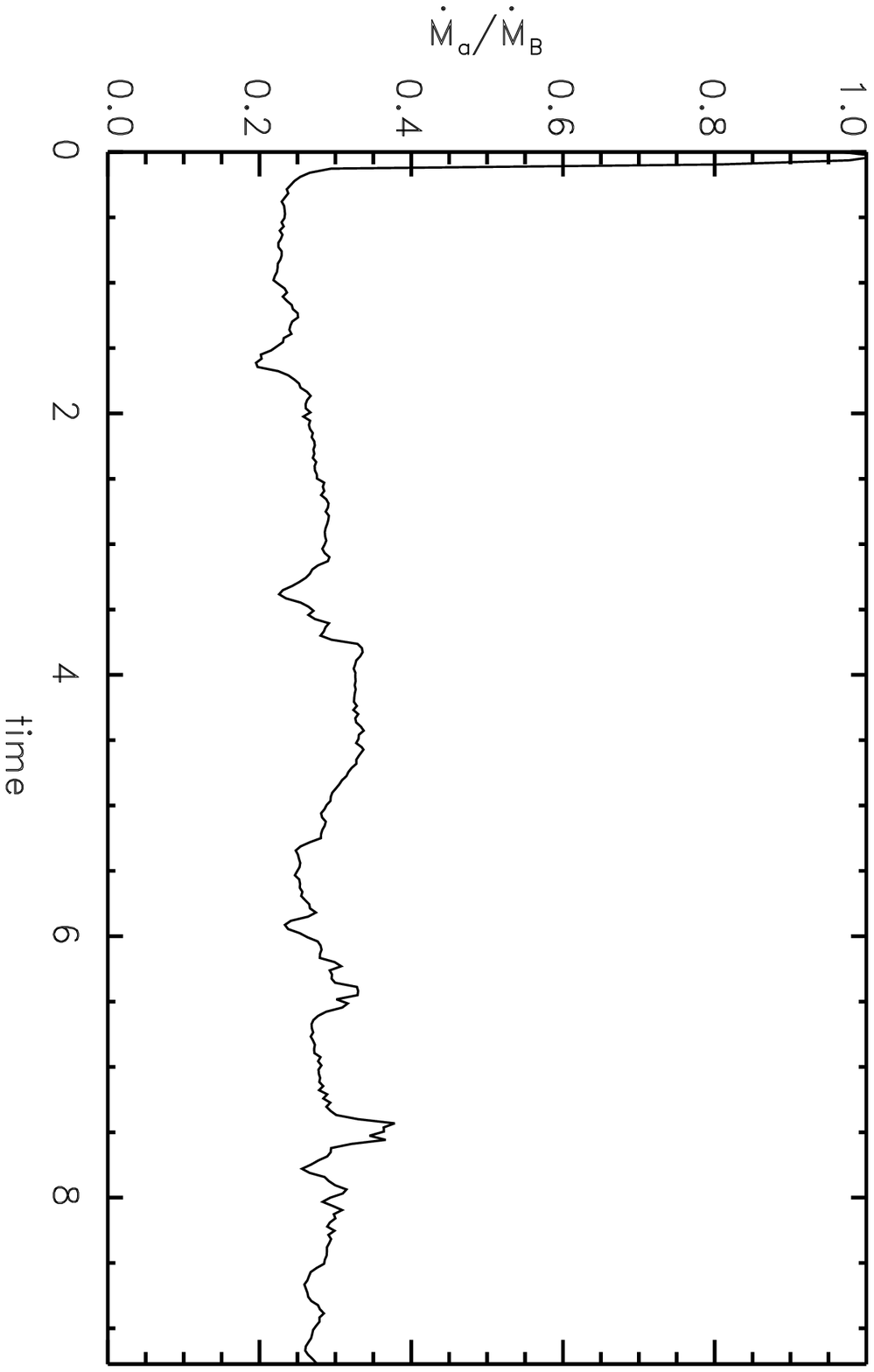}}
\end{picture}
\caption{ 
}
\end{figure}

\begin{figure}
\begin{picture}(180,330)
\put(-90,0){\includegraphics{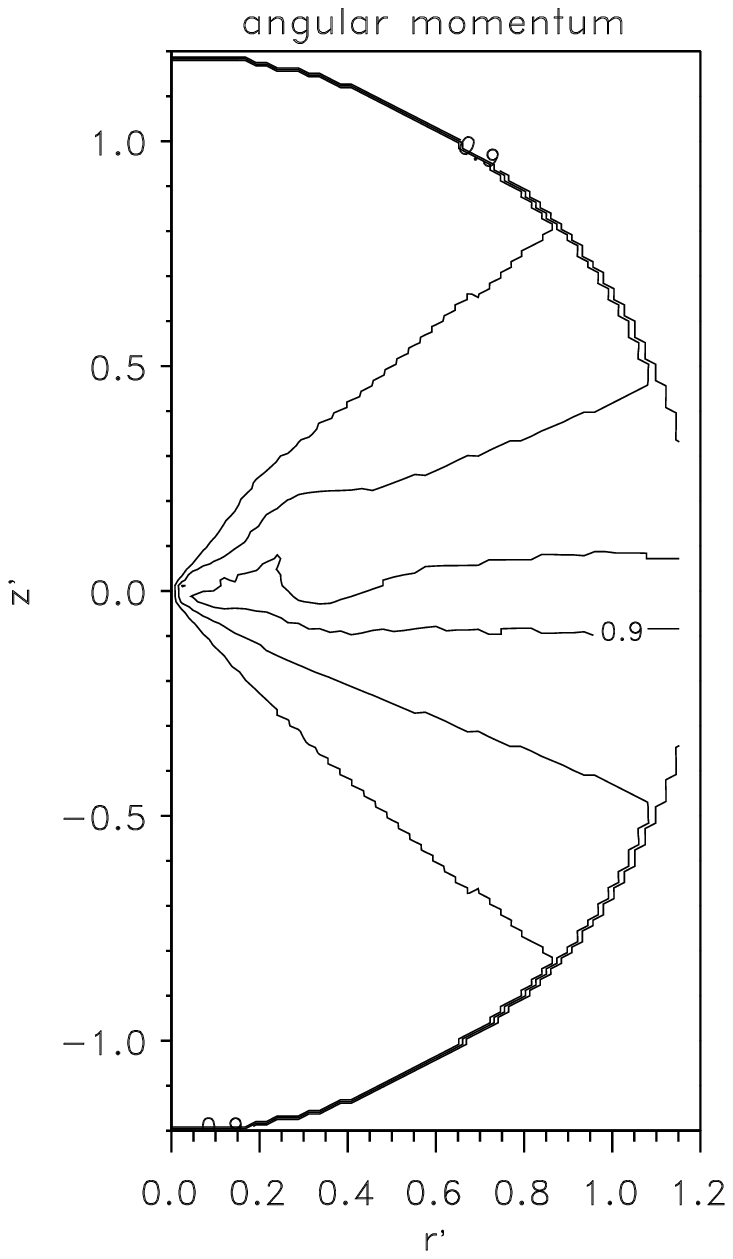}}

\put(80,0){\includegraphics{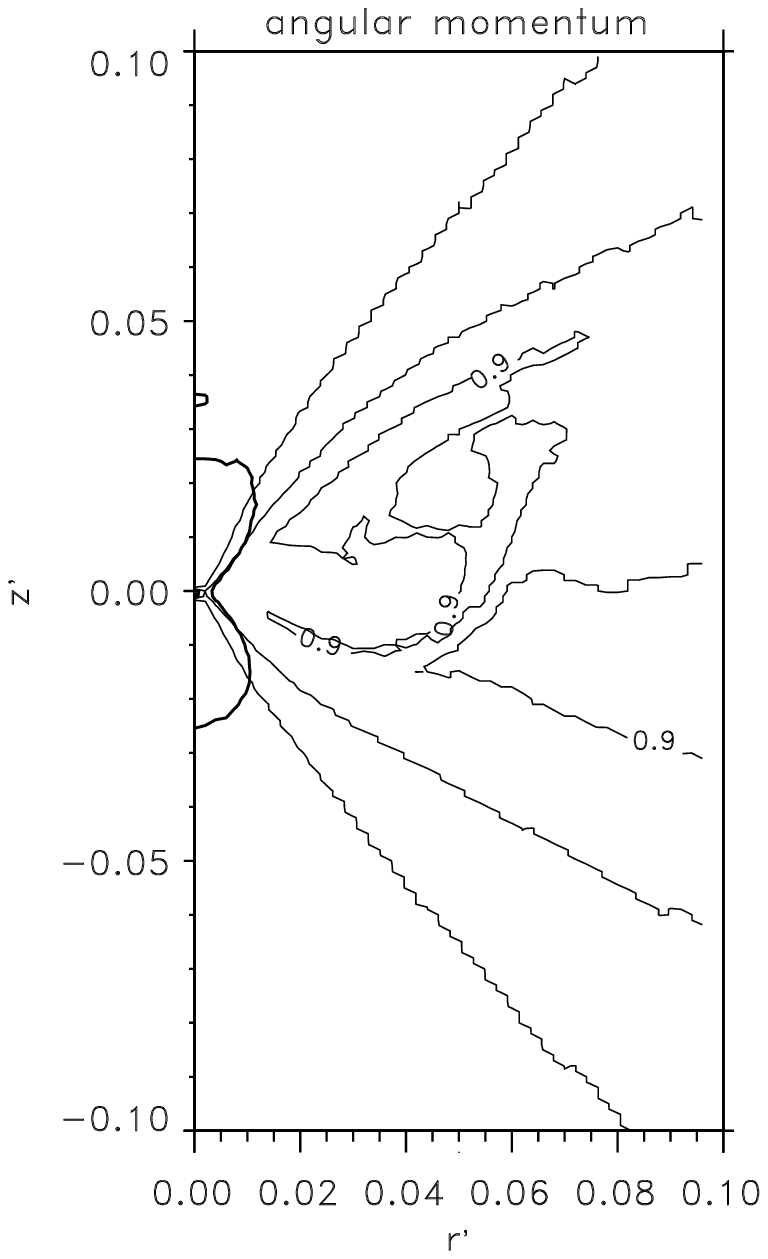}}

\end{picture}
\caption{ 
}
\end{figure}

\begin{figure}
\begin{picture}(180,400)
\put(0,0){\includegraphics{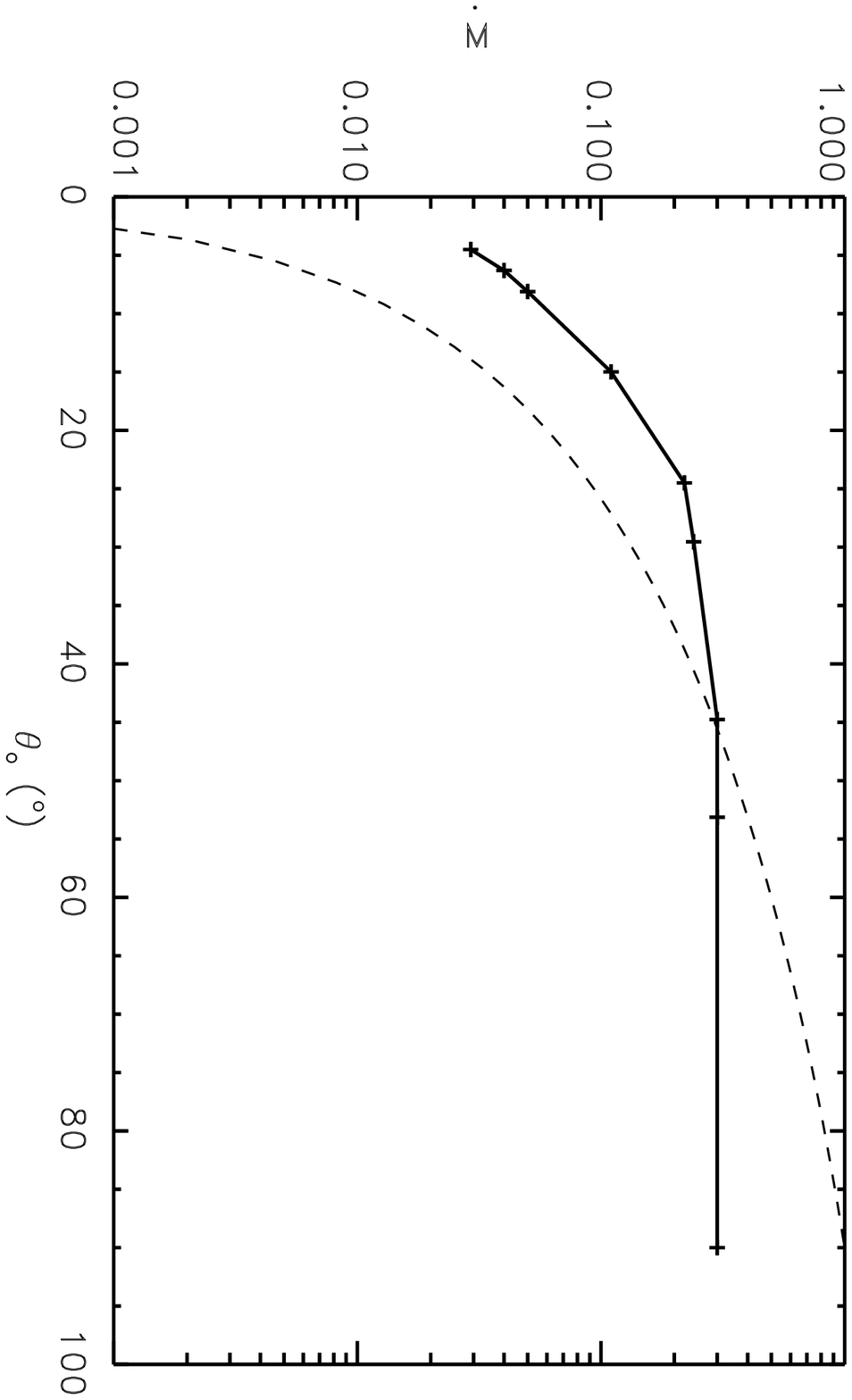}}
\end{picture}
\caption{ 
}
\end{figure}

\eject
\newpage

\end{document}